\documentclass[11pt,a4paper]{article}
\usepackage{jheppub}
\usepackage[utf8]{inputenc}
\usepackage{graphicx,fancybox,float,comment}

\usepackage{units}
\usepackage{subcaption}
\usepackage{multirow,array,arydshln}
\usepackage[dvipsnames]{xcolor}
\usepackage{braket,tensor}
\usepackage{amsmath,amssymb,amsfonts}
\usepackage{tikz-cd}

\usepackage{ulem}

\newcommand{\be}{\begin{equation}}
\newcommand{\ee}{\end{equation}}
\newcommand{\bea}{\begin{eqnarray}}
\newcommand{\eea}{\end{eqnarray}}
\newcommand{\bega}{\begin{gather}}
\newcommand{\eega}{\end{gather}}

\newcommand{\bi}{\begin{itemize}}
\newcommand{\ei}{\end{itemize}}
\newcommand{\ben}{\begin{enumerate}}
\newcommand{\een}{\end{enumerate}}
\newcommand{\bca}{\begin{cases}}
\newcommand{\eca}{\end{cases}}
\newcommand{\bln}{\begin{align}}
\newcommand{\eln}{\end{align}}
\newcommand{\bst}{\begin{split}}
\newcommand{\est}{\end{split}}
\def\ie{\begin{equation}\begin{aligned}}
\def\fe{\end{aligned}\end{equation}}
\newcommand{\bma}{\le(\begin{matrix}}
\newcommand{\ema}{\end{matrix}\ri)}

\def\le{\left}
\def\ri{\right}

\newcommand{\seq}{{s_{eq}}}
\newcommand{\sat}{{t_\mathrm{sat}}}
\newcommand{\sign}{{t_\mathrm{signal}}}

\global\interfootnotelinepenalty=1000

\graphicspath{{graphics}}

\title{The Page curve from the entanglement membrane}

\author[1]{Mike Blake}
\author[1, 2]{and Anthony P. Thompson}
\affiliation[1]{School of Mathematics, University of Bristol, \\ Woodland Road, Bristol BS8 1UG, U. K. }
\affiliation[2]{Quantum Engineering Center for Doctoral Training, University of Bristol, \\ Tyndall Avenue, Bristol BS8 1FD, U. K.}

\abstract{We study entanglement dynamics in toy models of black hole information built out of chaotic many-body quantum systems, by utilising a coarse-grained description of entanglement dynamics in such systems known as the `entanglement membrane'. We show that in these models the Page curve associated to the entropy of Hawking radiation arises from a transition in the entanglement membrane around the Page time, in an analogous manner to the change in quantum extremal surfaces that leads to the Page curve in semi-classical gravity. We also use the entanglement membrane prescription to study the Hayden-Preskill protocol, and demonstrate how information initially encoded in the black hole is rapidly transferred to the radiation around the Page time. Our results relate recent developments in black hole information to generic features of entanglement dynamics in chaotic many-body quantum systems.}

\begin{document}
\pdfoutput=1
\maketitle

\section{Introduction} 

\paragraph{} The past few years have seen significant new developments in understanding the fate of quantum information during the evaporation of a black hole \cite{Hawking:1975vcx, Hawking76}. In particular, it was appreciated by Page in the early 1990s that if black hole evaporation is unitary then the entanglement entropy of the emitted Hawking radiation should follow the Page curve \cite{Page_1993a,Page_1993b}, increasing until the black hole has approximately half evaporated before decreasing to zero. Remarkably, it was established in~\cite{penington2020entanglement, Almheiri_2019, Almheiri_2020} that the Page curve could be obtained from a computation in semi-classical gravity, by adapting the traditional quantum extremal surface prescription~\cite{Ryu_2006, Hubeny_2007, Engelhardt_2015} to include the contributions of so-called `islands'. The existence of such entanglement islands can be derived from the Euclidean gravitational path integral using the replica trick  \cite{penington2020replica, Almheiri_2020Replica}, where it is related to the existence of `replica wormhole' configuration in which different replicas of the gravitational system are coupled in a non-trivial manner\footnote{See \cite{Marolf_2021a, Marolf_2021b} for a Lorentzian interpretation of replica wormholes.}. Whilst these developments have led to significant progress, see e.g. \cite{Almheiri_2020HigherDim, Almheiri:2019yqk, Akers_2020, Hollowood:2020cou, Rozali_2020, Zhao_2020} for a selection of early work and~\cite{Almheiri_2021} for a review, they rely on complex details of the gravitational path integral. As such a clear physical understanding of entanglement islands and replica wormholes remains lacking. 

\paragraph{} A closely related question concerns the thermalization of a pure state $\ket{\psi}$ in a chaotic many-body quantum system. At late times such a state is expected to thermalize in the sense that measurements of local observables agree with those computed via a thermal density matrix $\rho$. Nevertheless, unitarity of quantum mechanics ensures that the system remains globally in a pure state, which has non-trivial consequences for the behaviour of non-local observables such as the von Neumann and Renyi entropies of a subregion $A$. For instance, for a globally pure state the entanglement entropies of $A$ and its complement must be equal, which lies at the heart of the original arguments for the Page curve~\cite{Page_1993b}. It is therefore natural to suggest that the recent insights into the unitarity of black hole evaporation are closely related to generic features of entanglement dynamics in chaotic many-body quantum systems.

\paragraph{} A particular microscopic connection between the Page curve and many-body quantum dynamics has been established in \cite{liu2019void, Liu_2021}. Specifically,~\cite{Liu_2021} studied simple models of evaporating and eternal black holes built out of chaotic quantum many-body systems. They found that various probes of entanglement dynamics in these models, including the Page curve and the Hayden-Preskill protocol \cite{Hayden_2007}, can be understood as arising microscopically from a dynamical mechanism known as `void formation' \cite{liu2019void}. Whilst these works represent a significant development, it is not straightforward to relate the microscopic process of void formation in such models to the recent computations in semi-classical gravity. With this motivation in mind, the purpose of this paper is to provide an alternative way of describing entanglement dynamics in similar toy models of evaporating and eternal black holes that more closely resembles the (quantum) extremal surface prescription of semi-classical gravity. 

\paragraph{} In order to do this, we will make use of an effective theory\footnote{A related general theory of information spreading in chaotic systems has been presented in \cite{Couch:2019zni}.} for the coarse-grained dynamics of von Neumann and Renyi entropies in chaotic many-body quantum systems with local interactions known as the `entanglement membrane'~\cite{Nahum_2017, jonay2018coarsegrained, Zhou_2019, PhysRevX.10.031066}. This description of entanglement dynamics can be explicitly derived for the case of 1+1 dimensional random quantum circuits\footnote{Numerical evidence in support of the entanglement membrane description for higher dimensional random circuits has recently been provided in \cite{sierant2023entanglement}.} with a large local Hilbert space dimension~\cite{Zhou_2019}, and holographic quantum field theories dual to classical gravity~\cite{Mezei_2018, Mezei_2020}\footnote{In this context the entanglement membrane exists in the space-time of the dual non-gravitational theory, and can be understood as the projection of the bulk HRT surface into the boundary.}. It has been conjectured to apply to a wide class of chaotic many-body quantum systems, although we note that at present the full extent of systems described by such an approach remains an important open question\footnote{In particular, it has been established that the simplest formulation of the entanglement membrane we will discuss in this paper does not describe the behaviour of higher ($n \geq 2)$ Renyi entropies for random circuits with diffusive conserved charges for which the entanglement entropy following a quench grows as$~\sqrt{t}$~\cite{PhysRevLett.122.250602, Huang_2020}. Other effects of conserved charges on entanglement dynamics have recently been studied in e.g. \cite{PhysRevB.107.045102, Majidy_2023}.}.  

\paragraph{} Within this effective description~\cite{Nahum_2017, jonay2018coarsegrained, Zhou_2019, PhysRevX.10.031066, Mezei_2018, Mezei_2020}, the entanglement entropies of a subregion of the chaotic many-body quantum system can be computed by minimising the entropy cost associated to a co-dimension one surface, the `entanglement membrane', in a slab of space-time. We will see that, when applied to models of evaporating and eternal black holes, the Page curve arises from a transition in the minimal membrane around the Page time, in an analogous manner to the change in the quantum extremal surface in semi-classical gravity. We will also use the entanglement membrane to probe the Hayden-Preskill protocol, and find that information encoded in the black hole is rapidly emitted around the Page time as anticipated in~\cite{Hayden_2007}.

\paragraph{} The remainder of this paper is organised as follows. In Section~\ref{sec:review} we briefly review the entanglement membrane description of entanglement dynamics in local chaotic many-body quantum systems. In Section~\ref{sec:evap0} we study a model of an evaporating black hole built out of a local chaotic many-body quantum system, and demonstrate how the entanglement membrane framework reproduces both the expected form of the Page curve and the Hayden-Preskill protocol in this model. In Section~\ref{sec:eternal} we apply the entanglement membrane to study these probes of entanglement dynamics in a model of an eternal black hole coupled to a bath. Our results are consistent with those derived explicitly in gravitational systems, and also the microscopic calculations performed in related toy models using void formation in~\cite{liu2019void, Liu_2021}. Finally, in Section~\ref{sec:discussion} we review important open questions, in particular commenting on the connection between the entanglement membrane and permutation degrees of freedom used in~\cite{Vardhan_2021} to connect the entanglement entropies of thermalized states to replica wormholes.

\section{Review of the entanglement membrane} 
\label{sec:review}

\paragraph{} We begin by reviewing a coarse-grained description of entanglement dynamics in certain chaotic many-body quantum systems based on an `entanglement membrane'~\cite{Nahum_2017, jonay2018coarsegrained, Zhou_2019, PhysRevX.10.031066}. The entanglement membrane is most easily motivated by results for random unitary circuits, but has a wider validity in more general chaotic systems including an appropriate limit of holographic quantum field theories~\cite{Mezei_2018, Mezei_2020}. This description can be derived analytically for `large-$q$' random circuits with a large local Hilbert space dimension~\cite{Zhou_2019, PhysRevX.10.031066}, and is consistent with numerical results for certain chaotic spin chains and Floquet models\footnote{In systems with conserved charges the simplest formulation of the entanglement membrane~\cite{jonay2018coarsegrained} we will discuss here assumes that after coarse-graining the system is in local equilibrium. We also note the even starting from homogeneous initial states, the presence of conserved charges can have non-trivial consequences for the behaviour of higher Renyi entropies. In particular, in random local quantum circuits with a diffusive conserved charge and finite Hilbert space dimension the higher Renyi entropies with $n > 1$ grow as $\sqrt{t}$ and are not described by the simplest formulation of the entanglement membrane we discuss here~\cite{PhysRevLett.122.250602, Huang_2020}.} \cite{Nahum_2017, jonay2018coarsegrained}.

\paragraph{} In particular we consider a homogeneous many-body quantum system, following a quench from a pure state $\ket{\psi(0)}$ at time $t=0$, with time evolution determined by chaotic unitary dynamics. At late times the system is expected to thermalise to an equilibrium configuration with entropy density $s_{eq}$. Our goal is to compute the entanglement entropy $S(A, t)$ of a subregion $A$ of the many-body quantum system during thermalization. In the entanglement membrane picture (valid on large lengthscales and timescales) the entropy $S(A,t)$ can be computed by minimising a functional describing the `entropy cost' associated to each choice of a co-dimension one membrane in a space-time slab of height $t$~\cite{Nahum_2017, jonay2018coarsegrained, Zhou_2019, PhysRevX.10.031066}.  
\begin{figure}
\begin{center}
\includegraphics[width=140mm]{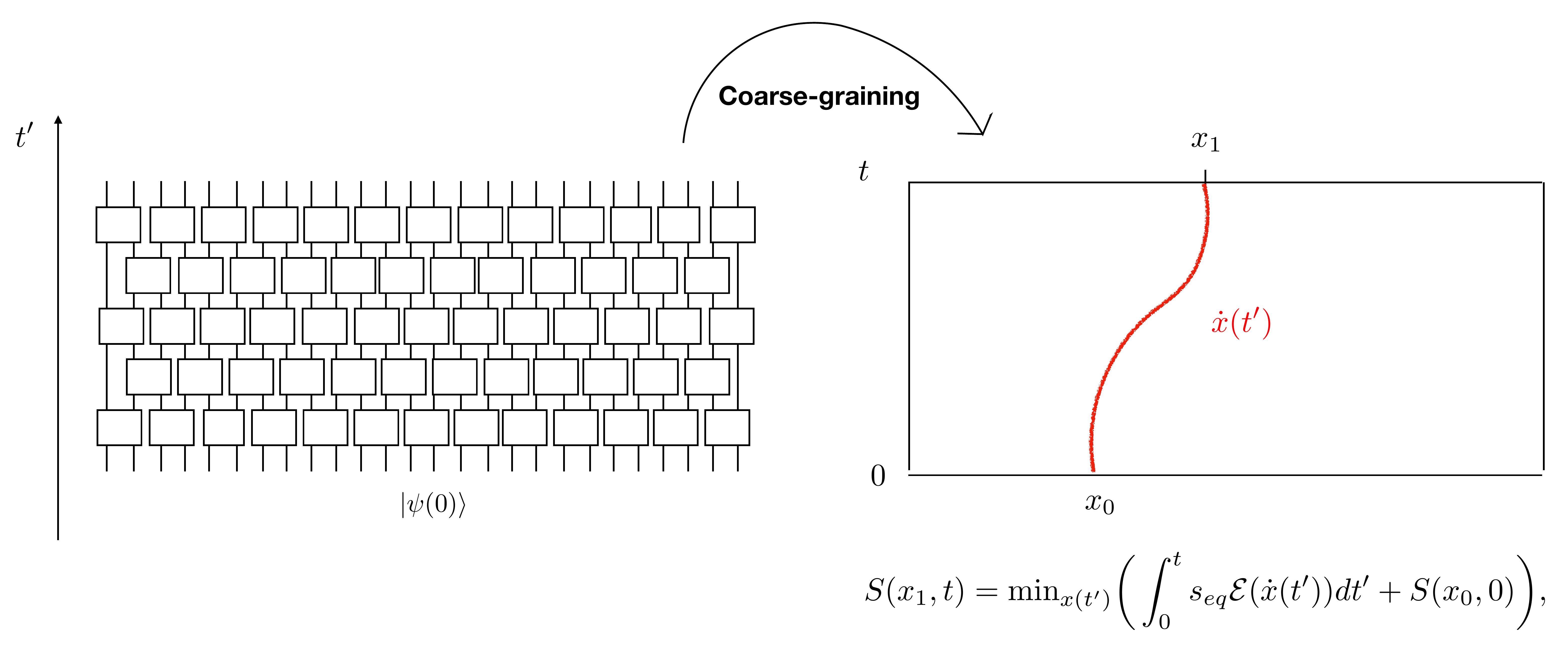}
\caption{Illustration of the entanglement membrane as a coarse-grained description of entanglement dynamics in random quantum circuits and chaotic many-body quantum systems.}
\label{fig:intro}
\end{center}
\end{figure}
\paragraph{} Whilst the entanglement membrane can be formulated in any dimension, the simplest setting in which to illustrate its use is an infinite one dimensional many-body quantum system with spatially homogeneous dynamics. We are interested in the entropy $S(x_1,t)$ across a cut in the system at position $x_1$, i.e. the entropy of the subregion $A = \{x : -\infty < x \leq x_1 \} $. To do this, one considers all curves $x(t')$ in a slab of spacetime $(x, t')$, where $t' \in [0, t]$ and $x(t')$ obeys the boundary conditions $x(t)=x_1$ and $x(0) = x_0$ with $x_0$ left free (see Figure~\ref{fig:intro}). The entropy can then be computed by minimising the following functional
\begin{equation}
S(x_1,t) = \mathrm{min}_{x(t')} \bigg( \int_0^{t} s_{eq} {\cal E}(\dot{x}(t')) dt' + S(x_0,0) \bigg), 
\label{entropy}
\end{equation}
where $s_{eq}$ is the equilibrium entropy density of the final state and $S(x_0,0)$ is the entropy of the region $\{x : -\infty < x \leq x_0 \}$ in the initial state $\ket{\psi(0)}$. The above prescription can naturally be modified to compute entanglement entropies of more/complicated subregions or in finite size systems.  For our purposes it will be particularly useful to note the following~\cite{jonay2018coarsegrained}:

\begin{itemize}
\item For a finite sized system (i.e. where $x \in [0, L]$) one must also consider membranes $x(t')$ that escape the space-time slab through either of the left or right hand edge of the system. For such membranes there is no contribution to the entropy cost from the initial state (i.e. one does not have the $S(x_0,0)$ term above).  
\item For a finite subregion $R = \{ x: x_1 \leq x \leq x_2 \}$ one includes membranes $x_L(t')$, $x_R(t')$ originating from each of the boundaries of $R$, i.e. satisfying $x_L(t) = x_1$ and $x_R(t) = x_2$. Such membranes may intersect (i.e. join) in the space-time slab\footnote{If the curves $x_L(t')$ and $x_R(t')$ intersect the initial state there is a contribution $S(R_0, 0)$ to the entanglement cost of the membrane from the entropy of the region $R_0$ enclosed by the curves in initial state.}.
\item For disjoint subregions $R = R_1 \cup R_2$ one includes curves emanating from the boundaries of both $R_1$ and $R_2$ at time $t$.
\end{itemize}
\paragraph{} The precise form of the line tension ${\cal E}(\dot x)$ depends both on the microscopic model and on the particular entanglement entropy being computed (e.g. the von Neumann $S_{vN}$ or second Renyi entropy $S_2$). Nevertheless, it can be argued that the line tension satisfies the following general properties~\cite{jonay2018coarsegrained, Zhou_2019}
\begin{equation}
{\cal E}(0) = v_E, \hspace{2.0cm} {\cal E}(v_B) = v_B, \hspace{2.0cm} {\cal E}'(v_B) = 1,
\label{equality}
\end{equation}
with $v_E$ the entanglement velocity of the entropy being computed and $v_B$ the butterfly velocity of the system of interest. Furthermore, we have the following inequalities
\begin{equation}
{\cal E}(v) \geq v, \hspace{2.0cm} {\cal E}''(v) \geq  0, \hspace{2.0cm} {\cal E}(v) \leq v_E + |v| \bigg(1 - \frac{v_E}{v_B} \bigg).
\label{inequality}
\end{equation}
We note that in minimising~\eqref{entropy} it is sufficient to consider only curves inside the lightcone $|\dot{x}| \leq v_B$. For a generic system in which these inequalities are satisfied strictly for $0 < v < v_B$ we have that both ${\cal E}(v)$, ${\cal E}'(v)$ are monotonically increasing functions in this domain. Finally we note that for 1+1 dimensional systems with homogeneous unitary dynamics then convexity of the line tension means that it is sufficient to only consider piece-wise linear curves \cite{jonay2018coarsegrained}.

\paragraph{} For certain microscopic models the line tension can be computed analytically. For instance if one considers a one dimensional chain with one spin of local Hilbert space dimension $q$ per unit length, then under random unitary dynamics a generic state $\ket{\psi(0)}$ thermalizes to an infinite temperature state with $s_{eq} = \ln q$. For a circuit built by applying random unitarities to bonds in a Poisson fashion (with rate 1), then as $q \to \infty$ one finds that the von Neumann entropy is governed by the line tension\footnote{In fact in the $q \to \infty$ limit the line tension is independent of the entropy being computed.}~\cite{jonay2018coarsegrained}
\begin{equation}
{\cal E}(v) = \frac{1}{2}(1 + v^2), 
\end{equation}
from which we can read off $v_E = 1/2$, $v_B = 1$ in such a model. In a second model with unitary dynamics built from a random brickwork circuit, then at large but finite $q$ (ignoring subleading $1/q$ corrections) one finds the second Renyi entropy is governed by the line tension\footnote{At strictly infinite $q$ the line tension ${\cal E}_2(v)$ becomes flat with ${\cal E}_2(v) =1$. This results in a highly degenerate minimal membrane problem. This degeneracy is resolved by working at large but finite $q$.}~\cite{jonay2018coarsegrained} 
\begin{equation}
{\cal E}_2(v) = \log_{q}  \frac{q^2 + 1}{q} + \frac{1 + v}{2} \log_{q} \frac{1 + v}{2} +  \frac{1 - v}{2} \log_{q} \frac{1 - v}{2}, 
\label{eps2}
\end{equation}
The line tension can also be computed for the von Neumann entropy of certain holographic quantum field theories~\cite{Mezei_2018, Mezei_2020}.
\paragraph{} Before applying the entanglement membrane to models of black hole evaporation, it will be useful to review how the entanglement membrane reproduces the expected results for the thermalization of a sub-region of a one dimensional system. In particular, consider the entanglement entropy $S(x,t)$ in a system of length $L$ starting from a product state (such that we can ignore any contribution to the entanglement cost from the initial state). At early times, the entropy is then minimised by a straight line dangling down vertically from the cut which gives $S(x,t) = s_{eq} {\cal E}(0) t = s_{eq} v_E t$. The entropy therefore initially increases linearly with time as expected. However, we can also consider membranes emanating from the cut at $(x,t)$ with slopes $v =  \pm v_B$.  For sufficiently large $t$ these will escape the system via the left/right edges of the space time slab and have a time-independent entropy cost. Assuming $x < L/2$ the cut escaping the left hand edge has the smaller entropy cost given by $S(x, t) = s_{eq} x$. For $t > x/v_E$ this is smaller than the entropy cost of the vertical membrane, and the entropy saturates at this thermalized value.

\section{Model of an evaporating black hole}
\label{sec:evap0}

\paragraph{} In the remainder of this paper we will apply the entanglement membrane to study entanglement dynamics in various models of black hole information. In this section we begin with a very crude model of information dynamics for an evaporating black hole. Whilst this model is over-simplified in many respects, it is instructive to see how the entanglement membrane reproduces both the expected form of the Page curve and the Hayden-Preskill protocol \cite{Hayden_2007} in this setting.

\paragraph{} In particular our model of an evaporating black hole consists of a homogeneous chaotic many-body quantum system of finite length $L$, with spatially local interactions. We will assume that the system is described by an entanglement line tension ${\cal E}(v)$, and thermalizes to an equilibrium configuration with uniform entropy density $s_{eq}$. We now divide the system into two parts. At each time $t$, we refer to the region $0 \leq x \leq x_B(t)$ where $x_B(t) = L - \alpha t$ as the black hole subsystem $B(t)$. Here $\alpha \ll v_E$ parameterises the rate at which the black hole evaporates, and $v_E = {\cal E}(0)$ is the entanglement velocity of the underlying chaotic many-body system\footnote{We caution the reader that in this section the entanglement velocity $v_E$ and butterfly velocity $v_B$ refer to the entanglement/butterfly velocities of the underlying local many-body system, and should not be viewed as those one would compute for a black hole in holography.}. The remaining degrees of freedom from $x_B(t) \leq x \leq L$ are referred to as the radiation subsystem $R(t)$. Note that the black hole and radiation systems are non-trivially coupled by the underlying local dynamics. A very similar model, in which the underlying microscopic system was a large-$q$ spin chain, was studied in Section II of \cite{Liu_2021} using void formation. 
\begin{figure}
\begin{center}
\includegraphics[width=80mm]{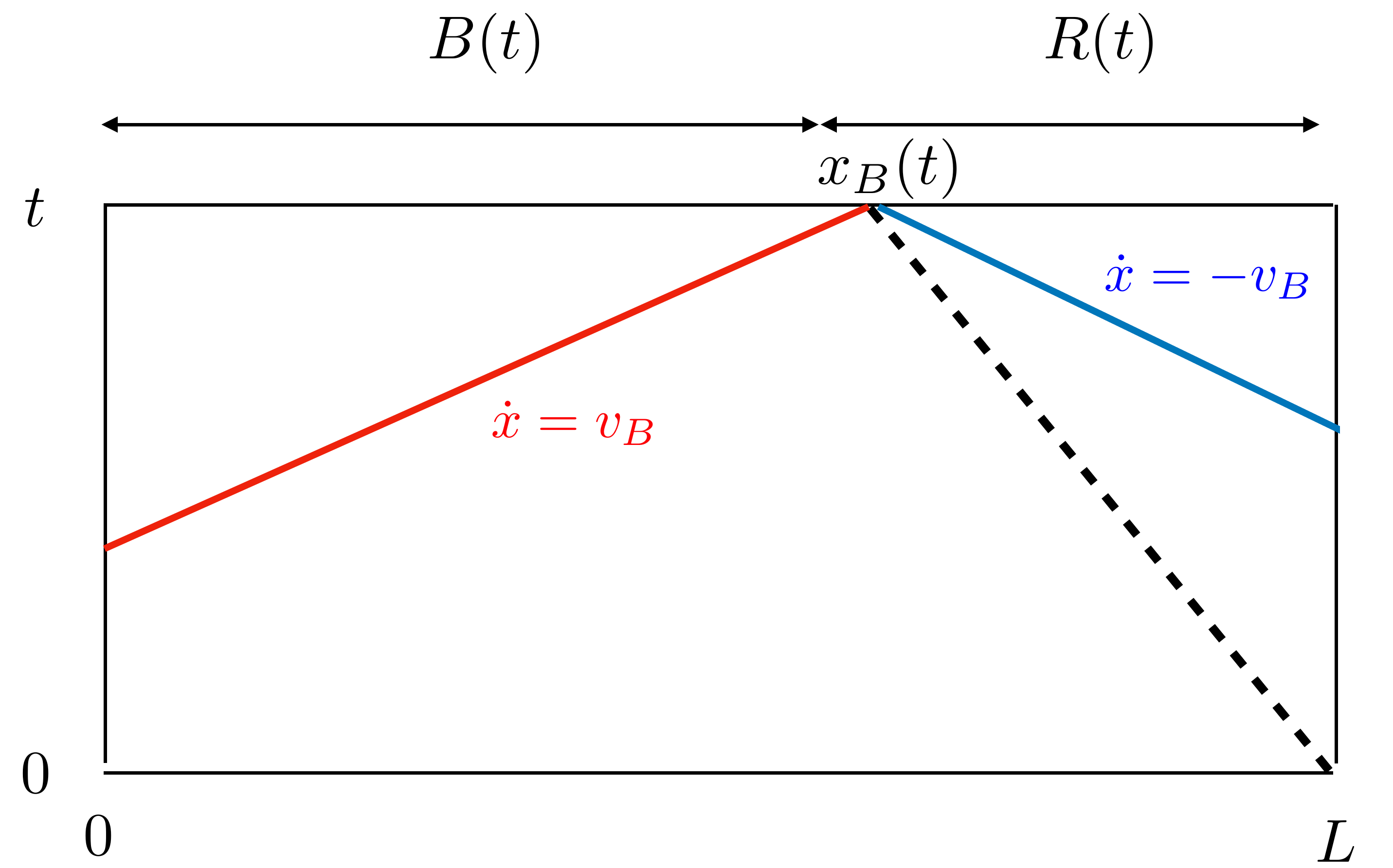}
\caption{Entanglement membranes responsible for the Page curve in a toy model of an evaporating black hole.}
\label{fig:islands0}
\end{center}
\end{figure}
\paragraph{} We are interested in computing the entanglement entropy of the black hole region $B(t)$. For simplicity we will assume that the initial state of the system at $t=0$ is a product state, such that the initial entropy profile across a cut is trivial. Then for $\alpha \ll v_E$ the behaviour of the entanglement entropy $S(B(t))$ can be understood from a competition between two minimal surfaces of slopes $|\dot{x}| = v_B$. In particular, the relevant membranes consist of:
\begin{enumerate}
\item A curve from $x_B(t)$ to the right boundary at $x=L$ with slope $\dot{x} = -v_B$. This has entropy $\alpha s_{eq} t$.
\item A curve from $x_B(t)$ to the left boundary at $x=0$ with slope $\dot{x} =  v_B$. This has entropy $s_{eq} (L - \alpha t) = S_{BH} - \alpha s_{eq} t$ where we have introduced $S_{BH} = s_{eq} L$. 
\end{enumerate}
These surfaces are illustrated in Figure~\ref{fig:islands0}. Note that for this set up a vertical membrane emanating from $x_B(t)$ has entropy cost $ v_E s_{eq} t \gg \alpha s_{eq} t$ and hence is never minimal for a slowly evaporating black hole. 
 \paragraph{} Precisely the same membranes are also relevant for the computation of the radiation subregion $R(t)$. We therefore find the entropies:
\begin{align}
S(B(t)) = S(R(t)) &= \alpha s_{eq}  t, && t \leq t_P, \nonumber \\
&= S_{BH} - \alpha s_{eq} t, && t \geq t_P,
\end{align} 
which we can identify as the Page curve with Page time $t_{P} = S_{BH}/(2 \alpha s_{eq})$. The sharp change in the behaviour of the entropy around the Page time arises from a change in the minimal membrane at $t = t_P$, analogous to the transition between bulk extremal surfaces responsible for the Page curve in the gravitational computation of entanglement entropy. Furthermore, we note that in the above computation we have not needed to specify the precise entropy being computed -  the above results apply to both von Neumann and Renyi entropies, provided that the entanglement membrane prescription is valid for the quantity of interest.

\subsection{Information transfer from an evaporating black hole}
\label{sec:evap}
\paragraph{} As a more refined probe of information dynamics during black hole evaporation, we now turn to study the Hayden-Preskill protocol \cite{Hayden_2007}. In particular, we are interested in understanding how information originally inside a (young) black hole is transferred to its Hawking radiation during evaporation. 

\paragraph{} In order to study this in the context of our toy model, we introduce an additional reference system $Q$ of length $p \ll L$, consisting of the same degrees of freedom as those of the black hole/radiation subsystem $B(t) \cup R(t)$. The reference system $Q$ is not coupled by any interaction to the evaporating black hole system, however we assume it has non-trivial entanglement with the black hole degrees of freedom at $t=0$. In particular we assume the reference system $Q$ is entangled with an identical subsystem $P \subset B(0)$ (i.e. of length $p \ll L$) such that the entropy density in $P$ at $t=0$ is given by $s_{eq}$ \footnote{Here we are choosing the entropy density of $P$ in the initial state to be the same as that of the state the system ultimately thermalizes to. Other choices for the initial entropy density can also be considered and give qualitatively similar results.}. The remaining degrees of freedom of $B(0)$ are taken to be initially in a product state. We are then interested in how the information in $Q$ initially encoded in the black hole subsystem through this entanglement is spread to the radiation subsystem as the black hole evaporates. According to the Hayden-Preskill protocol, this information is expected to be rapidly transferred to the radiation around the Page time \cite{Hayden_2007}. 

\begin{figure}
\begin{center}
\includegraphics[width=90mm]{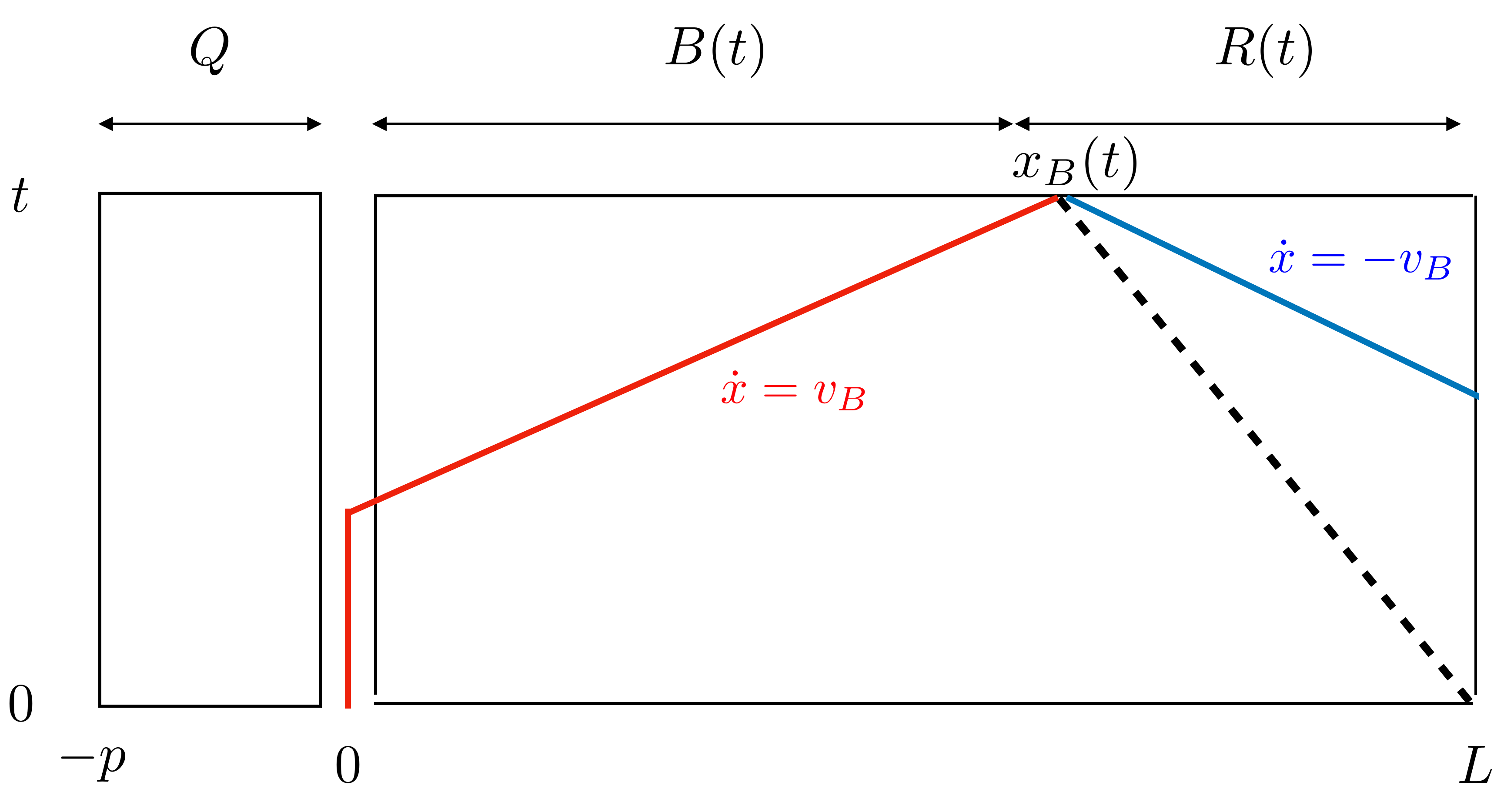}
\caption{Extremal membranes relevant for $S(R(t))$ for a young black hole entangled with a reference system.}
\label{fig:islands6}
\end{center}
\end{figure}

\paragraph{} In order to study this, we will use the entanglement membrane to compute the (Renyi or von Neumann) mutual information between $Q$ and the subsystem $R(t)$ 
\begin{eqnarray}
I(Q,R(t)) &=& S(Q) + S(R(t)) - S(Q  \cup R(t)), \nonumber \\
&=&  S_Q + S(R(t)) - S(B(t)), \;\;\;\;\;\;\;\;\;\;\;\; S_Q = s_{eq} p ,
\label{mutualQR}
\end{eqnarray}
where we have used the overall system $Q \cup  R(t) \cup B(t)$ is in a pure state. An analogous formula for $I(Q, B(t))$ can be obtained by interchanging $R(t)$ and $B(t)$ in \eqref{mutualQR}. 

\paragraph{} In order to proceed, it is necessary to adapt the entanglement membrane prescription to apply in the presence of a reference system $Q$. To do this, it is useful to consider the subsystem $Q$ to lie to the left hand side of the black hole-radiation system, i.e. in the region $-p \leq x < 0$. Since there are no interactions coupling $Q$ to the region $B(t) \cup R(t)$, we now consider applying the entanglement membrane in a geometry in which there is a thin gap in the space-time slab at $x=0$ (see Figure~\ref{fig:islands6}). We further will assume that any piece of the entanglement membrane in this gap in the space-time slab gives rise to no additional cost to the entropy (i.e. ${\cal E}(v) = 0$ in the gap), except for a possible contribution from the initial state. Such a prescription is natural from the `minimal cut' picture of the entanglement membrane in random circuits in which the entropy cost is associated with the number of non-trivial gates in the circuit cut by the membrane~\cite{Nahum_2017}. 

\begin{figure}
\begin{center}
\includegraphics[width=90mm]{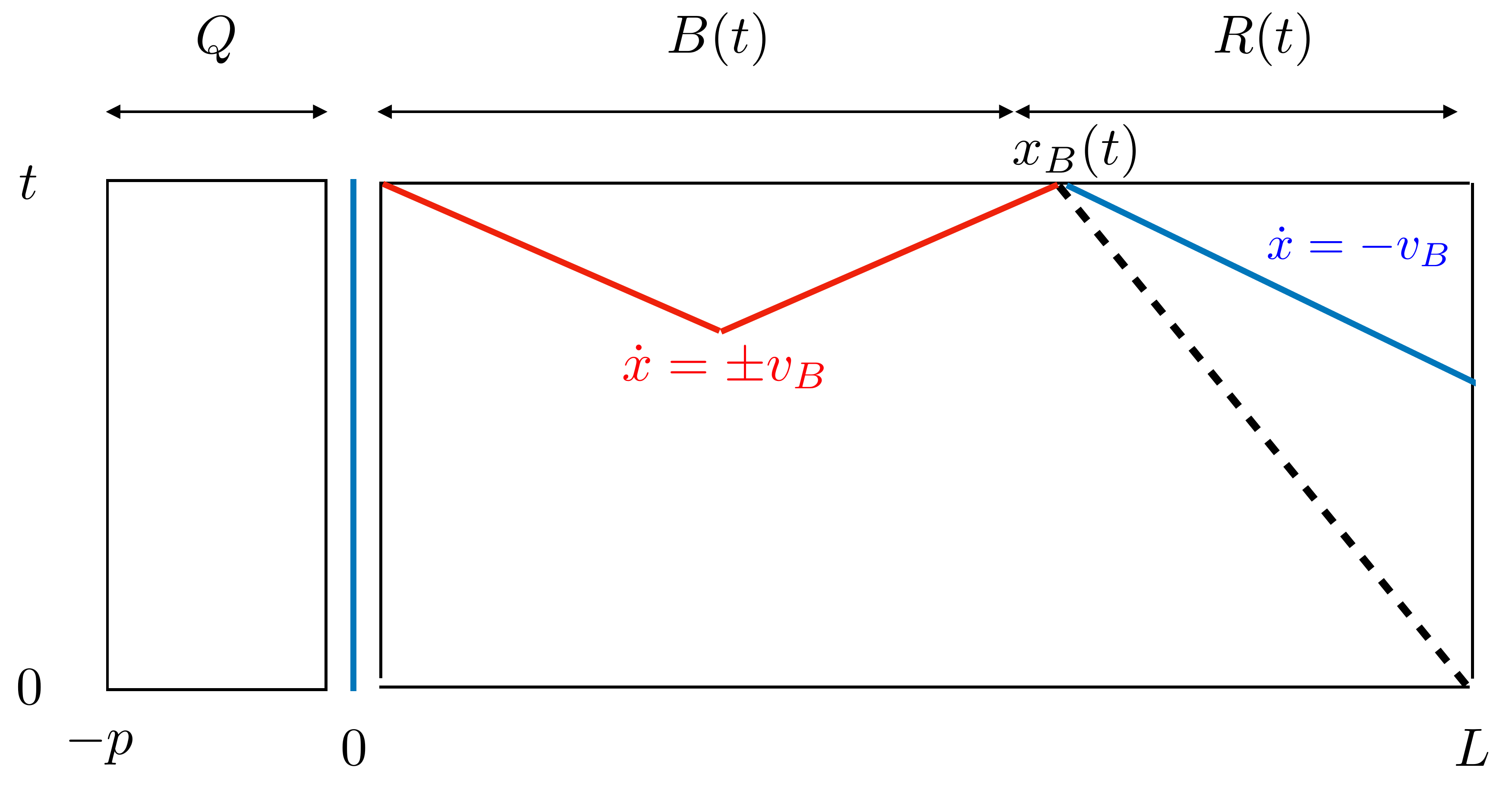}
\caption{Extremal membranes for $S(B(t))$ for a young black hole entangled with a reference system.}
\label{fig:islands7}
\end{center}
\end{figure}

\paragraph{} The computation of $S(R(t))$ can then be understood from the following (candidate) minimal surfaces (illustrated in Figure~\ref{fig:islands6}).
\begin{enumerate} 
\item A surface that emanates from $x_B(t)$ and crosses to $x=L$ with slope $\dot{x} = - v_B$.  This has entropy $\alpha s_{eq} t$. 
\item A surface that emanates from $x_B(t)$ and crosses to $x=0$ with $\dot{x} = v_B$, before dropping down vertically to $t=0$ through the gap between $Q$ and $B(t)$. This gives a contribution to the entropy of $S_{BH} - \alpha s_{eq} t$ from the surface passing through $B(t)$ and a contribution $S_Q = s_{eq} p$ from the membrane cutting the initial state at $x=0$.  
\end{enumerate}
 The entropy is given by the minimum of these two surfaces, i.e.
 \begin{align}
 S(R(t)) &=\alpha s_{eq} t, && t \leq t_P + \Delta t, \nonumber \\
 &= S_Q + S_{BH}  -  \alpha s_{eq} t, &&  t \geq t_P + \Delta t, 
\end{align} 
where $t_P = L/2 \alpha$ is the Page time and $\Delta t = p/2 \alpha$. We can compute $S(B(t))$ in an entirely analogous way, with the relevant surfaces now shown in Fig.~\ref{fig:islands7}. Computing the entropy cost of each membrane and choosing the smaller we find 
 \begin{align}
 S(B(t)) &= S_Q + \alpha s_{eq} t , && t \leq t_P - \Delta t, \nonumber \\
 &=  S_{BH} - \alpha s_{eq} t, && t \geq t_P - \Delta t.
\end{align} 
The above results precisely match equations (2.25) and (2.26) of \cite{Liu_2021}, when applied to the microscopic model of a large-$q$ spin chain considered there (for comparison in the notation of \cite{Liu_2021} one has $\alpha = 1$, $s_{eq} = \ln q$, $S_{BH} = k$). They imply the mutual informations have the form:
\begin{align}
 I(Q, R(t)) &= 0, &&t \leq t_P - \Delta t, \nonumber \\
 &= S_Q - S_{BH}+ 2 \alpha s_{eq} t, &&  t_P - \Delta t \leq t \leq t_P + \Delta t, \nonumber \\
 &= 2 S_Q, && t \geq t_P + \Delta t, 
 \end{align}
and similarly 
\begin{align}
 I(Q, B(t)) &= 2 S_Q, && t \leq t_P - \Delta t, \nonumber \\
 &= S_Q + S_{BH} - 2 \alpha s_{eq} t , &&  t_P - \Delta t \leq t \leq t_P + \Delta t,\nonumber \\
 &= 0, && t \geq t_P + \Delta t. 
 \end{align}
\begin{figure}
\begin{center}
\includegraphics[width=70mm]{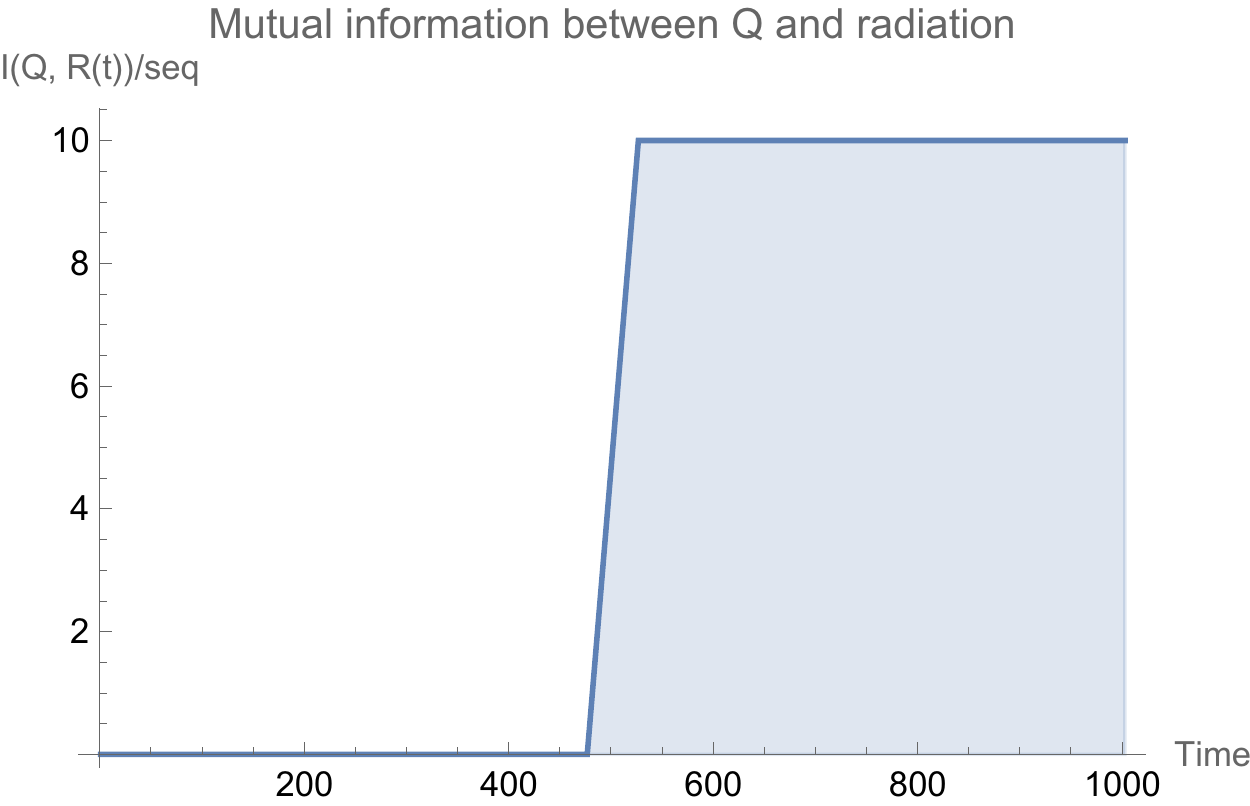}
\includegraphics[width=70mm]{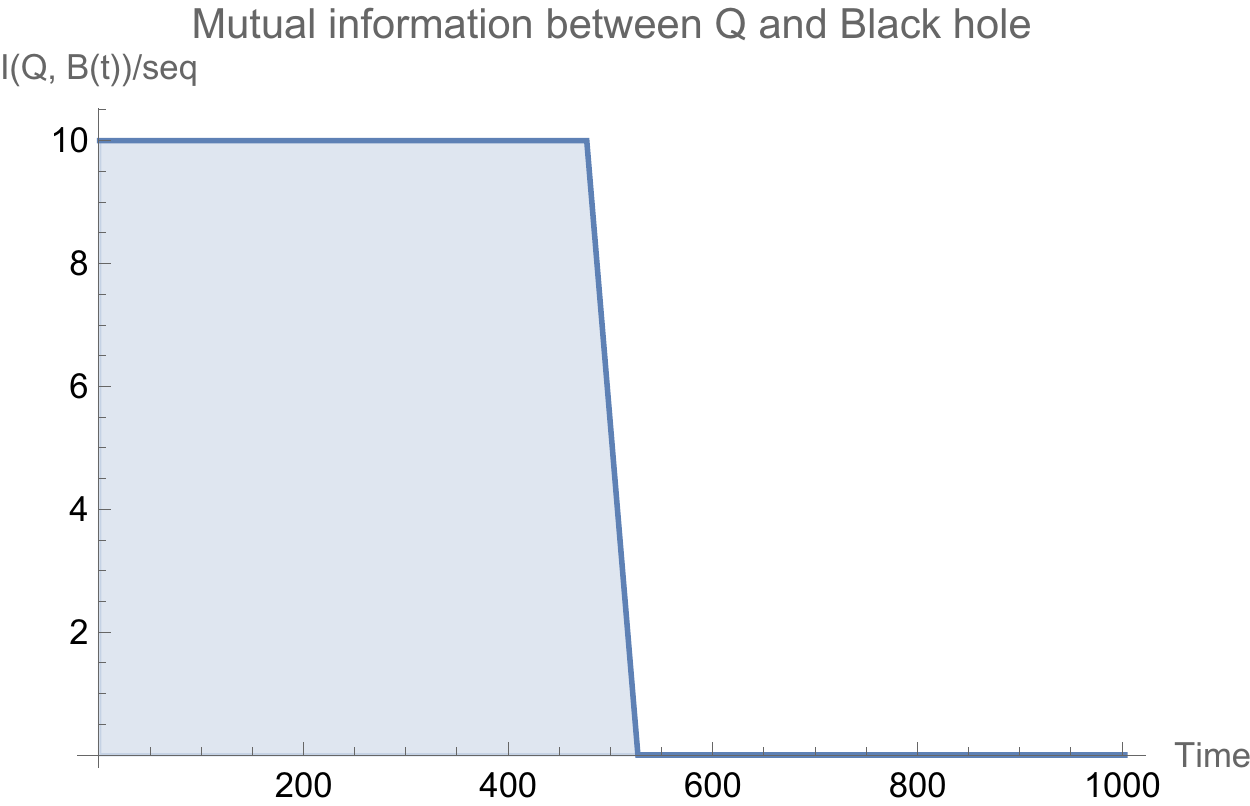}
\caption{Mutual information between reference $Q$ and radiation/black hole subsystems in a model of an evaporating black hole. Here the parameters have been set to $p = 5$, $L = 100$, $\alpha = 0.1$, $t_P = 500 $.}
\label{mutualfig}
\end{center}
\end{figure}
These results show how the information about the reference system $Q$ initially contained in the black hole subsystem is rapidly transferred to the radiation subsystem around the Page time in a time of order $\Delta t$ (see Figure~\ref{mutualfig}). As mentioned above, this is the expected behaviour implied by the arguments of Hayden and Preskill for a young\footnote{Here young refers to the fact the information is encoded in the black hole significantly before the Page time.} black hole~\cite{Hayden_2007}. Here we see how, in the context of our toy models, this transfer of information is associated to the transition between different minimal membranes in the computations of $S(B(t))$ and $S(R(t))$. 

\section{Model of an eternal black hole}
\label{sec:eternal}

\paragraph{} We now consider information dynamics using the entanglement membrane in the context of a toy model of an eternal black hole coupled to a bath. The Page curve in this setting was originally studied in \cite{Almheiri:2019yqk}, and our model is closely related to a toy dynamical model of an eternal black hole studied in \cite{Liu_2021}. In particular, we wish to apply the entanglement membrane to study information dynamics in an inhomogeneous system consisting of two black hole regions $B_1, B_2$ and two bath/radiation regions $R_1, R_2$ in the following one dimensional system\footnote{The equilibrium entropy of the bath $s_{eq} L$ should be viewed as parametrically larger than the black hole entropy $S_{BH} = \tilde{s}_{eq} a$. In all cases we discuss we take $S_{BH}$ sufficiently large that the Page time is parametrically larger than the other timescales that appear in this section.}:
\begin{enumerate}
\item  A bath region $R_1$ in the interval $[ -(L+a), -a]$ and $R_2$ in $[a, L + a]$, with $a \ll L$. The bath regions consist of a local chaotic many-body system with equilibrium entropy density $s_{eq}$ and entanglement line tension ${\cal E}(v)$.
\item Two black hole regions $B_1$ in the interval $[-a, 0]$ and $B_2$ from $[0, a]$ consisting of a second local chaotic many-body system with a parametrically larger equilibrium entropy density $\tilde{s}_{eq} \gg s_{eq}$ and an entanglement line tension $\tilde{{\cal E}}(v)$. 
\end{enumerate}
A concrete example to have in mind is a random circuit model, in which the bath regions $R_1, R_2$ consist of spins with a $q$-dimensional Hilbert space with equilibrium entropy density $s_{eq} = \ln q$ and the black hole regions $B_1, B_2$ consist of spin with a $q^N$-dimensional Hilbert space ($N \gg1)$ and equilibrium entropy density $\tilde{s}_{eq} = N \ln q$. 
\paragraph{}We assume that the left(right)-hand regions $R_1 (R_2)$ and $B_1(B_2)$ are coupled locally by the dynamics. However there are no interactions between the left hand system $R_1, B_1$ and the right hand system $R_2, B_2$. The geometry within which we can apply the entanglement membrane is then illustrated in e.g. Figure~\ref{fig:islands3}, where we emphasise that the entanglement line tension and equilibrium density is different in the black hole and bath regions. Since we are interested in modelling an eternal black hole, we assume that at $t=0$ the black hole regions $B_1$ and $B_2$ are in a thermofield double state (or maximally entangled if a random circuit model) with equilibrium density $\tilde{s}_{eq}$. In contrast for the purposes of the discussion in the main text we will assume that the bath regions $R_1, R_2$ are initially in a product state at $t=0$\footnote{It is possible to also study the entanglement dynamics when $R_1$ and $R_2$ initially have non-trivial entanglement, as was done in the model studied in Section III of~\cite{Liu_2021} and we discuss in~Appendix~\ref{app:mutual}.}. 

\paragraph{} Note that in contrast to the related examples studied in~\cite{Almheiri:2019yqk, Liu_2021}, where the black hole regions were 0+1 dimensional systems, our black hole regions $B_1, B_2$ have a finite width $a \ll L$ and local dynamics in the black hole region. These assumptions are necessary in order to be able to apply the entanglement membrane in these regions. However we will see that the limit $a \to 0$ with $S_{BH} = \tilde{s}_{eq} a$ fixed can be freely taken in our results for both the Page curve and Hayden-Preskill protocol below - these do not depend on the details of the black hole dynamics, but only the total entropy of the black hole region.
\subsection*{Page curve for an eternal black hole} 
\begin{figure}
\begin{center}
\includegraphics[width=120mm]{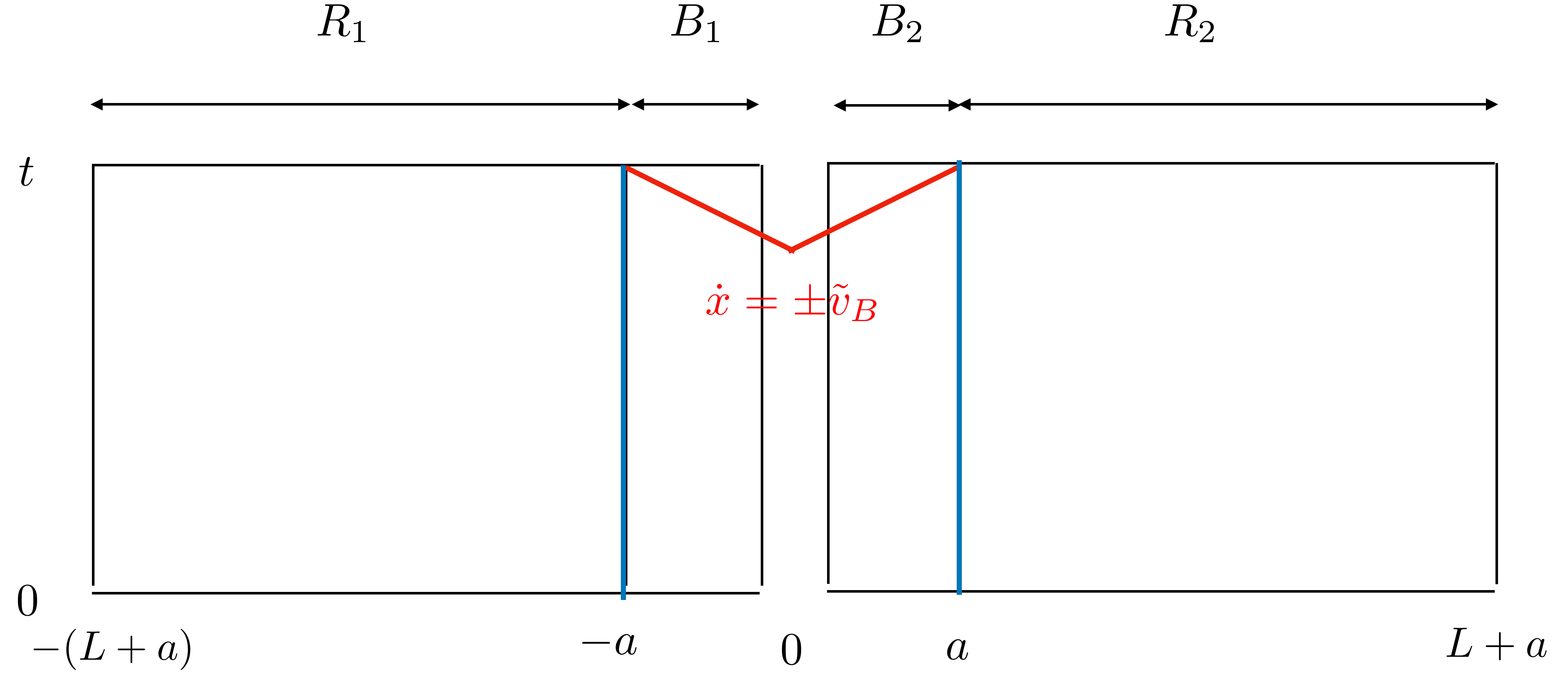}
\caption{Surfaces contributing to the entropy of the bath $S(R, t)$ in our toy model of an eternal black hole. }
\label{fig:islands3}
\end{center}
\end{figure}
\paragraph{} We begin by computing the entanglement entropy of the bath region $R = R_1 \cup R_2$ in our model of an eternal black hole. At time $t=0$ this is zero since we have assumed that $R_1$ and $R_2$ are initially in a pure state. The relevant entanglement membranes for computing this entropy are shown in Figure~\ref{fig:islands3}. They consist of:
\begin{enumerate}
\item A membrane which consists of two disconnected vertical lines near $x=\pm a$. These lines should be viewed as lying infinitessimally inside the radiation regions $R_1$ and $R_2$, as vertical lines inside the black hole regions would have a larger entropy cost due to the parametrically larger entropy density inside the black hole $\tilde{s}_{eq} \gg s_{eq}$\footnote{Here we are assuming the entanglement velocities for the bath and black hole regions are comparable, as is the case for a random circuit model built of spin-$q$ (spin-$q^N$) degrees of freedom in the bath (black hole) regions.}. The entropy cost of this membrane is then $S(R, t) = 2 s_{eq} {\cal E}(0) t = 2 s_{eq} v_E t $, with $v_E$ the entanglement velocity of the bath region. 
\item A closed surface inside the black holes formed from straight lines of velocities $\dot{x} = \pm \tilde{v}_B$, with $\tilde{v}_B$ the butterfly velocity in the black hole region\footnote{This should not be viewed as the butterfly velocity of a holographic theory.}. This contributes an entropy $S = 2 S_{BH}$, where $S_{BH} = \tilde{s}_{eq} a$.  
\end{enumerate}
Picking the entropy of the minimal membrane we then find
\begin{align}
S(R,t ) &= 2 s_{eq} v_E t, && t \leq t_P, \nonumber \\
 &= 2 S_{BH},  && t \geq t_P,
\label{eternalBHentropy}
\end{align}
where the Page time $t_P = S_{BH}/(s_{eq} v_E)$ is when the two membranes exchange dominance. Note that for the concrete case of a large $q$ random quantum circuit model discussed above, then we have $s_{eq} = \ln q$ and $S_{BH} = k \ln q$ where $k = a N$. The Page time would then evaluate to $t_P = k/v_E$, where the entanglement velocity depends on the entropy being computed and the precise circuit. Our results for $S_2$ for a large-$q$ brickwork circuit (for which $v_E = 1$) then coincide with those obtained in~\cite{Liu_2021} for a related dynamical model of an eternal black hole using void formation. 

\paragraph{} Before moving on to discuss the Hayden-Preskill protocol in our model of an eternal black hole, we note it was argued in \cite{Liu_2021} that void formation leads to the generation of mutual information $I(R_1, R_2, t)$ between the radiation regions around the Page time. In Appendix~\ref{app:mutual} we show how this physics arises from the point of view of the entanglement membrane. 

 \subsection{Information transfer from inside an eternal black hole}
 
 \begin{figure}
\begin{center}
\includegraphics[width=90mm]{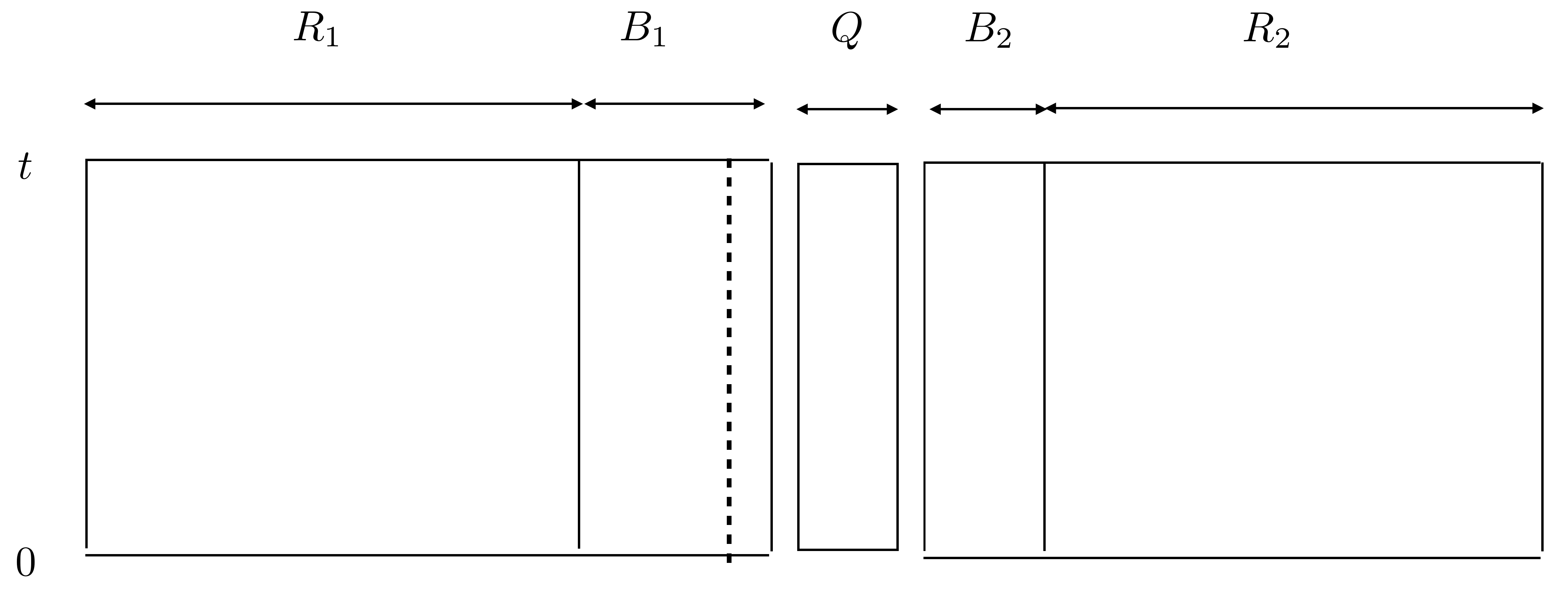}
\caption{Illustration of the space-time set up for studying information transfer in our model of an eternal black hole. The dotted line is to indicate that the region $B_1$ has been extended by additional degrees of freedom $P \subset B_1$.}
\label{fig:islands8}
\end{center}
\end{figure}

\paragraph{} We now turn to study the Hayden-Preskill protocol in our model of an eternal black hole. We begin by studying how information initially encoded inside the black hole is transferred to the radiation/bath regions over time. Motivated by the analysis of~\cite{Liu_2021}, we add a small number of additional degrees of freedom to the black hole $B_1$ by expanding its width to $a + p$, where $p \ll a$. We will denote these additional black-hole degrees of freedom as $P \subset B_1$. We then consider the initial state of $P$ to be entangled with a reference system $Q$, such that $S(Q) = S(P) = \tilde{s}_{eq} p$ at $t=0$. Otherwise, the set up is the same as in the previous subsection - i.e. the remaining spins in $B_1$ are initially in a thermofield double state with a second black hole region $B_2$ of width $a$ with entropy density $\tilde{s}_{eq}$, and the radiation regions $R_1 (R_2)$, are taken to initially be in product states. 

\paragraph{} A diagrammatic representation of the space time slab in which we will apply the entanglement membrane is shown in Figure~\ref{fig:islands8}. Note that the system consists of three subsystems $(R_1, B_1)$, $(R_2, B_2)$ and $Q$ which are decoupled (through interactions) but have non-trivial entanglement between them. We can consider arranging these subsystems in any order - for the purpose of drawing entanglement membranes we find it convenient to place $Q$ in between the left and right black hole/bath subsystems as shown in the figures. 
\paragraph{} As in Section~\ref{sec:evap} we are interested in computing the mutual informations $I(Q, B, t)$ and $I(Q, R, t)$, where $B = B_1 \cup B_2$ and $R = R_1 \cup R_2$.  We have
\begin{eqnarray}
I(Q,R,t) &=& S(Q) + S(R, t) - S(Q  \cup R, t), \nonumber \\
&=&  S_Q + S(R, t) - S(B, t), \hspace{3.0cm} S_Q = \tilde{s}_{eq} p
\label{mutual}
\end{eqnarray}
and an identical expression for $I(Q,B,t)$ by exchanging $R$ and $B$. We again just need to compute the entropies $S(R, t)$ and $S(B, t)$.
\begin{figure}
\begin{center}
\includegraphics[width=90mm]{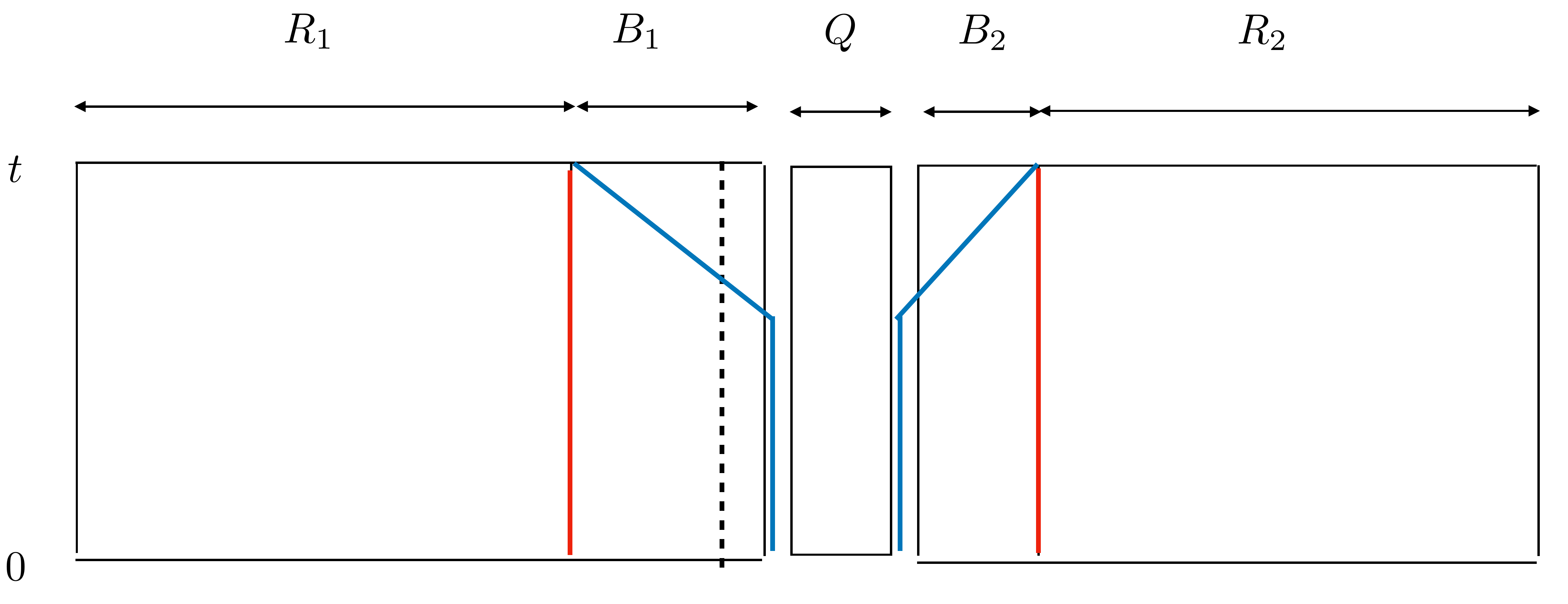}
\caption{Extremal surfaces for entropy of bath regions $S(R, t)$ in our model of an eternal black hole.}
\label{fig:islands9}
\end{center}
\end{figure}
\paragraph{}The behaviour of $S(R, t)$ can be understood from the entanglement membranes in Fig.~\ref{fig:islands9}, consisting of:
\begin{enumerate}
\item Two vertical surfaces at the edges of each bath each contributing an entropy $s_{eq} v_E t$. As in the previous section these surfaces should be viewed as lying infinitessimally inside the bath regions. 
\item A surface that cuts across the black hole regions ${B}_1$, ${B}_2$ with velocities $\dot{x} = \pm \tilde{v}_B$ and then drops down to $t=0$ through the gaps in the space-time slab around the reference system $Q$. This surface contributes an entropy $2 S_{BH} + 2 S_{Q}$. Note one factor of $S_Q$ is arising due to the addition of the extra degrees to freedom to $B_1$, and the second factor of $S_Q$ from the fact the membrane encloses the region $Q$ in the initial state. 
\end{enumerate}
Picking the minimal surface gives
 \begin{align}
 S(R, t) &= 2 s_{eq} v_E t, && t \leq t_P + \Delta t, \nonumber \\
 &= 2 S_{BH} + 2 S_Q, &&  t \geq t_P + \Delta t,
\end{align} 
where $t_P = S_{BH}/(s_{eq} v_E)$ is the Page time and $\Delta t  = S_Q/(s_{eq} v_E) \ll t_P$. 
\paragraph{} Likewise we find the behaviour of $S(B, t)$ from the following entanglement membranes (see Fig.~\ref{fig:islands10}):
\begin{figure}
\begin{center}
\includegraphics[width=90mm]{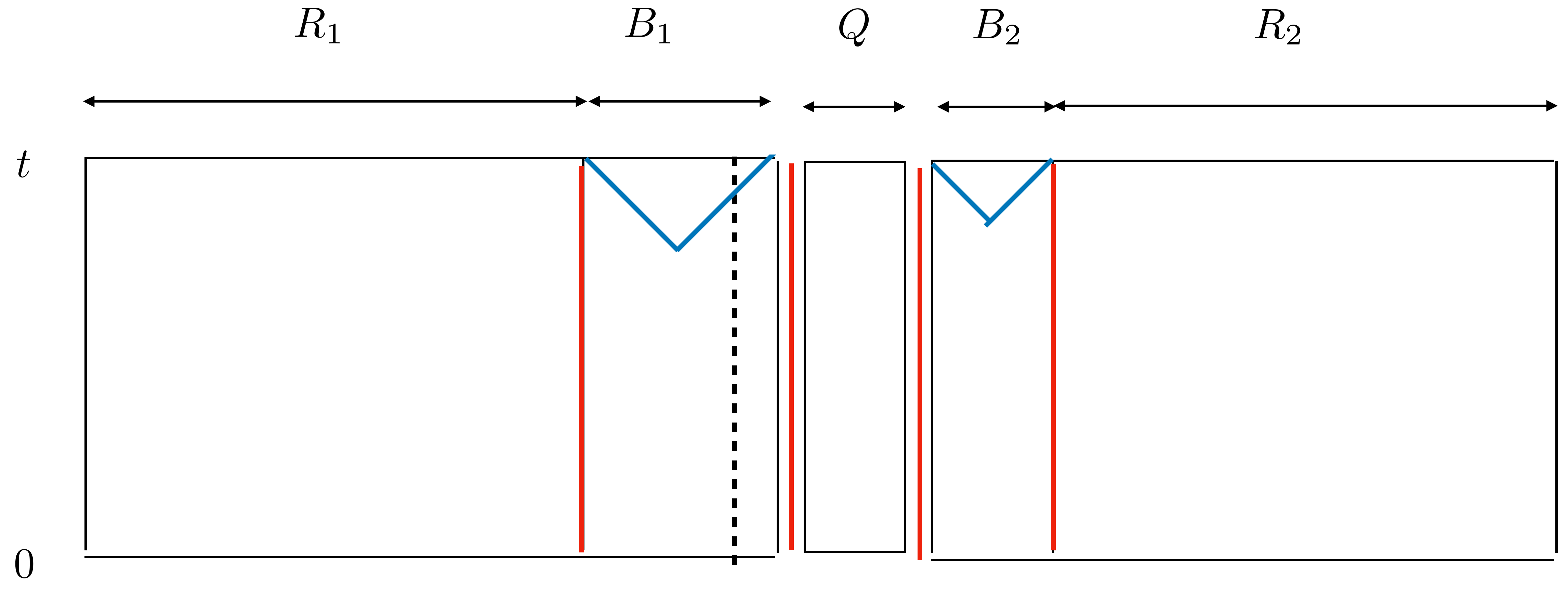}
\caption{Extremal surfaces for the entropy of black hole region $S(B, t)$ in our model an eternal black hole.}
\label{fig:islands10}
\end{center}
\end{figure}
\begin{enumerate}
\item A disjoint surface consisting of four vertical surfaces near the edges of $B_1$ and $B_2$. These surfaces should be viewed as two vertical surfaces infinitesimally inside the bath regions, each contributing an entropy $s_{eq} v_E t$, and two interior surfaces that run down to the initial state through the gaps in the space time slab. The total entropy is then $2 s_{eq} v_E t + S_{Q}$, with the factor of $S_Q$ coming from the region enclosed by the membrane in the initial state. 
\item The join of two surfaces that cut across the interiors of the regions ${B}_1$ and ${B}_2$ with $\dot{x} = \pm \tilde{v}_B$ contributing a total entropy $2 S_{BH} + S_{Q}$. 
\end{enumerate}
We therefore find
 \begin{eqnarray}
 S(B, t) &=& 2 s_{eq} v_E t + S_{Q},  \hspace{3.7cm} t \leq t_P,\nonumber \\
 &=& 2 S_{BH} + S_{Q} , \hspace{4.0cm}  t \geq t_P.
\end{eqnarray} 
From which we obtain the mutual informations
\begin{eqnarray}
 I(Q, R,t) &=& 0, \hspace{6.2cm} t \leq t_P, \nonumber \\
 &=& 2 s_{eq} v_E t - 2 S_{BH}, \hspace{3.7cm}   t_P \leq t \leq t_P + \Delta t,  \nonumber \\
 &=& 2 S_Q, \hspace{5.7 cm} t \geq t_P + \Delta t,
 \end{eqnarray}
and similarly 
\begin{eqnarray}
 I(Q, B, t) &=& 2 S_Q, \hspace{5.7cm} t \leq t_P, \nonumber \\
 &=& 2 S_Q  + 2 S_{BH} -  2 s_{eq} v_E t,  \hspace{2.6cm}   t_P  \leq t \leq t_P + \Delta t, \nonumber \\
 &=& 0, \hspace{6.2 cm} t \geq t_P + \Delta t. 
  \end{eqnarray} 
From which we see that the information in $Q$ is transferred rapidly (in a time $\Delta t$) from the black hole to the bath/radiation around the Page time. Note our results precisely match those of~\cite{Liu_2021} when applied to the case of $S_2$ for a brickwork large-$q$ quantum circuit. This can be seen by noting that in the context of the model in~\cite{Liu_2021} one has $s_{eq} =\ln q$, $v_E = 1$, $S_{BH} = k \ln q$ and hence $t_P = k$, $\Delta t = p$. 

 \subsection{Information transfer from outside an eternal black hole}\label{InfOut}
 \begin{figure}
\begin{center}
\includegraphics[width=90mm]{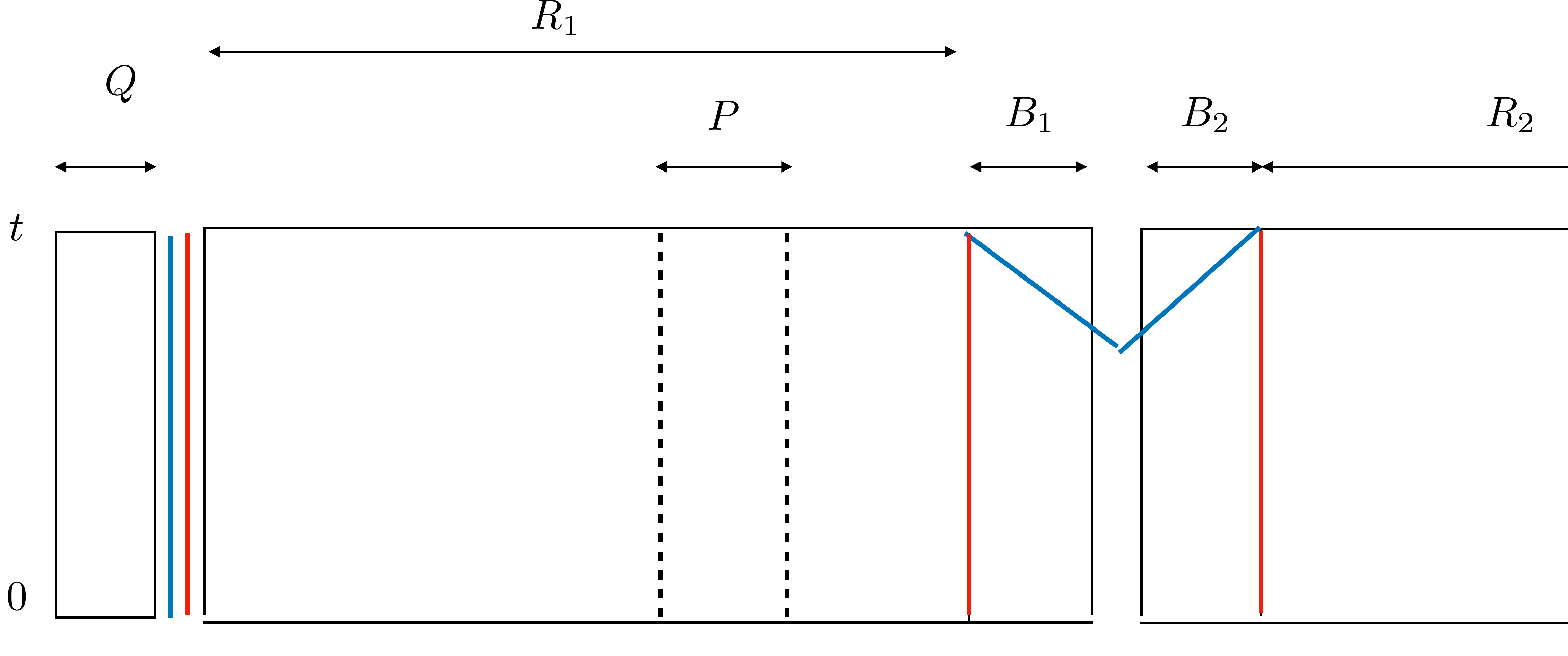}
\caption{Simple diagrams contributing to entropy of bath $S(R, t)$ during transfer of information from outside an eternal black hole.}
\label{fig:islands12}
\end{center}
\end{figure}
\paragraph{} We now turn to consider what happens if the information contained in the reference system $Q$ is entangled with bath degrees of freedom outside the eternal black hole. We will see that the coupling between the bath and the black hole means that some of this information is rapidly transferred to the black hole region, and the precise details of this information transfer have some model dependence.  However in the cases we study we find this information is rapidly recovered around the Page time, consistent with the expectations of~\cite{Hayden_2007}. 

\paragraph{}  To this end we consider extending the bath region $R_1$ by length $p$, and again refer to these new degrees of freedom as subsystem $P$.  We will take the region $P$ to be located a distance $l$ away from the interface between $R_1$ and the black hole region $B_1$. We then again consider an initial state in which the new bath degrees of freedom in $P$ are entangled with a reference system $Q$, such that initially $S(P) = S(Q) = s_{eq} p$ and $P$ has uniform entropy density $s_{eq}$. As before, we take the bath degrees of freedom outside of $P$ to initially be in a product state, and the black hole regions $B_1, B_2$ to be initially in a thermofield double state with density $\tilde{s}_{eq}$.
\paragraph{} We are interested in computing the mutual information $I(Q, R, t)$. We will see that the early time behaviour, describing the transfer of information to the black hole region, is sensitive to two particular timescales. In particular let us first define $t_{\mathrm{signal}} = (l + p)/v_B$, with $v_B$ the butterfly velocity in the bath region. This corresponds to the time required for a signal emitted from the edge of $B_1$ to reach the left hand side of the region $P$. Further, we define $t_{\mathrm{sat}} = l/v_E$. We will see the early time behaviour of the mutual information differs depending on which of these timescales is larger.

\subsubsection{$t_{\mathrm{sat}} < t_{\mathrm{signal}}$} %
 \begin{figure}
\begin{center}
\includegraphics[width=90mm]{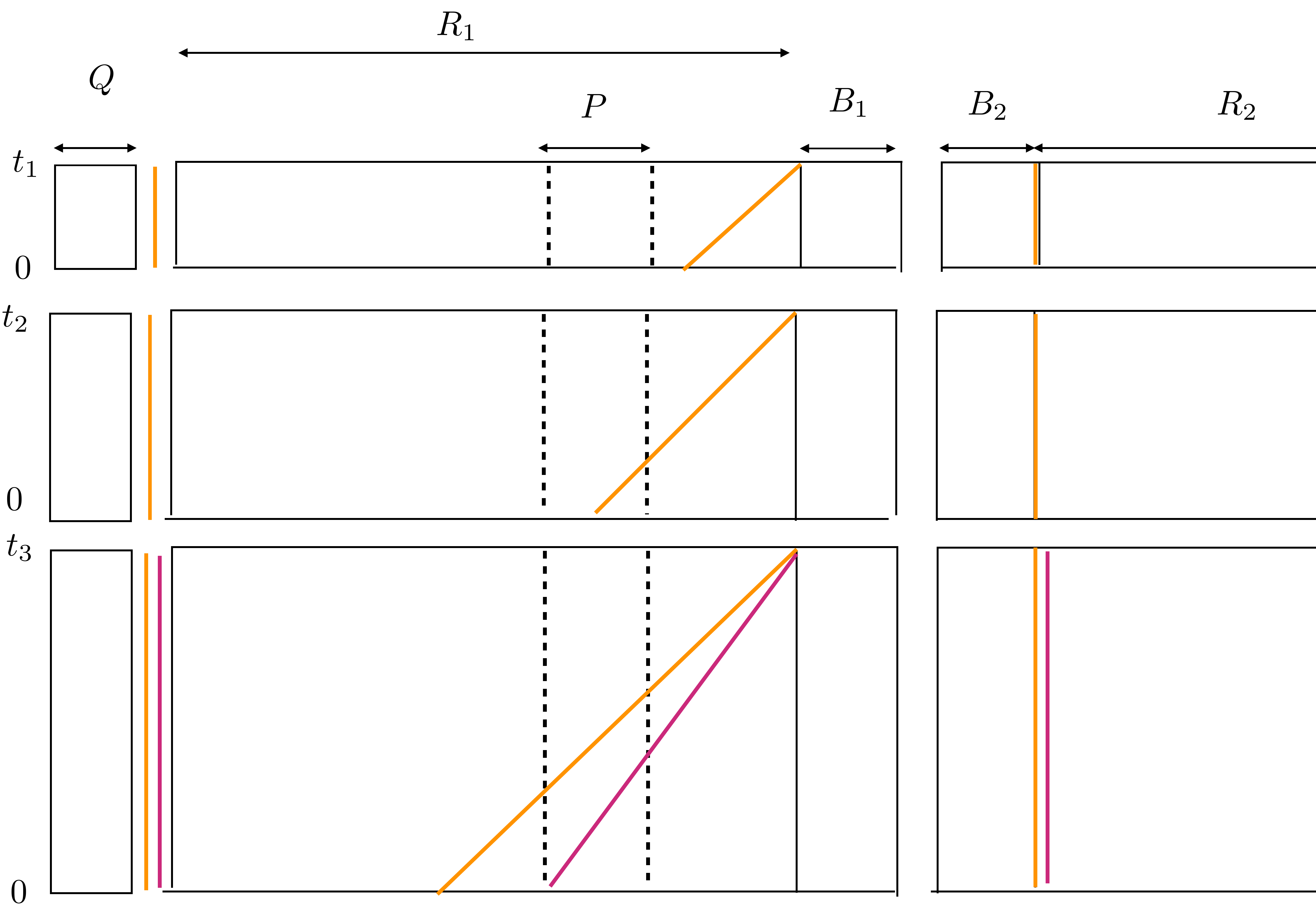}
\caption{Candidate membranes for $S(R,t)$ at intermediate times during the Hayden-Preskill protocol for our model of an eternal black hole. The membranes are shown for three qualitatively different times.}
\label{fig:islands13}
\end{center}
\end{figure}
\begin{figure}
\begin{center}
\includegraphics[width=60mm]{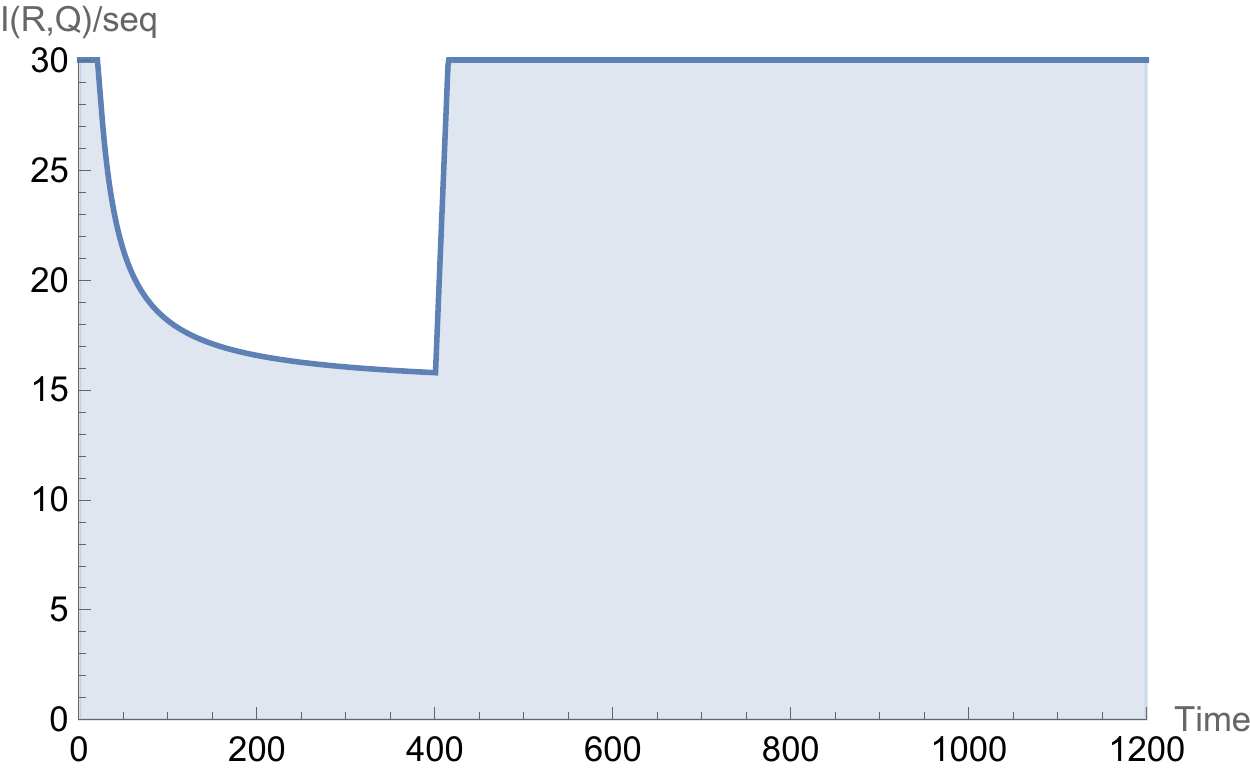}
\includegraphics[width=60mm]{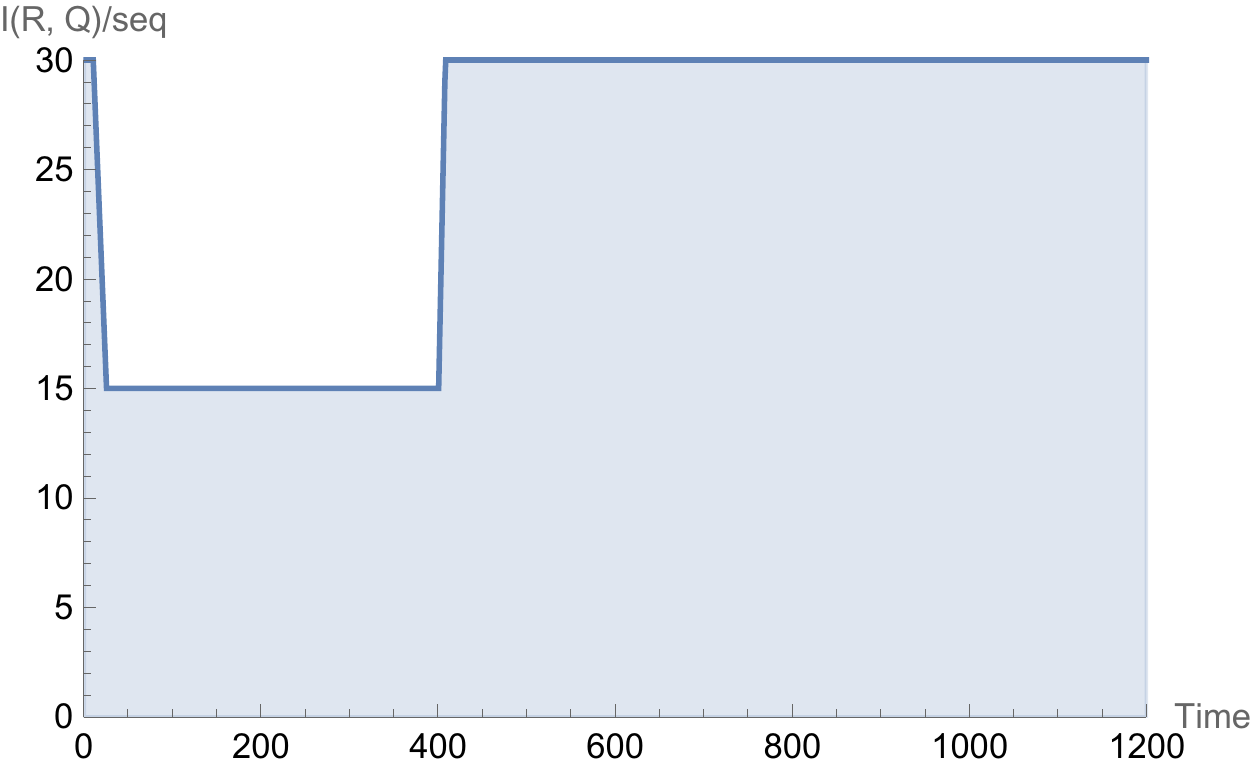}
\caption{Mutual information between reference system $Q$ and bath $R$  during the Hayden-Preskill protocol in models of an eternal black hole built from random large $q$ circuits. The left hand plot shows $I_{vN}(Q, R,t)/s_{eq}$ in a Poisson circuit for $l=10, p =15,  t_P = 400$. The right hand plot shows $I_{2}(Q,R,t)/s_{eq}$ in a brickwork circuit for $l=10, p =15,  t_P = 400$.}
\label{fig:mutual}
\end{center}
\end{figure}

\paragraph{} First, we will consider the case when $t_{\mathrm{sat}} < t_{\mathrm{signal}}$, i.e. we have $l/v_E < (l+p)/v_B$. In this case there are four different types of membrane to consider that are relevant for the entropy $S(R,t)$. The first two are shown in Figure~\ref{fig:islands12}, and are analogous to cases we have studied previously. They contribute $S_Q  + 2 s_{eq} v_E t$ (shown in red, which gives the early time behaviour of $S(R, t)$) and $S_Q + 2 S_{BH}$ (shown in blue, which gives the late time saturation value of $S(R, t)$).
\paragraph{} However, at intermediate time-scales there are additional membranes that become minimal, due to the inhomogeneous entropy profile of the initial state due to the  entanglement between $P$ and $Q$. The first intermediate time membrane is shown as the orange membrane in Figure~\ref{fig:islands13}, shown for three qualitatively different times.  This membrane corresponds to a straight line with gradient $\dot{x} = v_{B}$ from the interface between $B_1$ and $R_1$ into the bath region. This results in a cut along the bottom of our spacetime a distance $v_B t$ away the black hole region. The contribution of this curve is different depending on the value of $t$, and we find it becomes minimal for $l/v_E < t <  (l+p)/v_B$. In particular, focusing for simplicity on the contributions to the entropy from the part of the membrane in the left hand system $(R_1, B_1)$ we have:
\begin{enumerate}
\item For $t < l/v_B$ this part of the membrane contributes an entropy $s_{eq} v_B t $ from the slope part plus an entropy $S_{Q}$ from the fact the membrane cuts the initial state to the right of $P$. This gives a total entropy $S_{Q} + s_{eq} v_B t > S_{Q} + s_{eq} v_E t$ and hence this cut is not minimal. 
\item For $l/v_B < t < (l+ p)/v_B$ we again have a contribution $s_{eq} v_B t$ from the slope, but now the cut across the initial state lies in the region $P$. The entropy across the cut is now given by $S_{Q} - s_{eq}(v_B t-l)$ and hence the total contribution of this surface is $S_{Q} + s_{eq} l$ which is $t$ independent. For $t > l/v_E$ this is less than $S_{Q} + s_{eq} v_E t$ coming from a vertical line and hence this becomes part of the minimal surface. 
\item For $t > (l + p)/v_B$ there is a contribution $s_{eq} v_B t$ from the slope, but now the cut lies to the left of region $P$ and there is no contribution to the entropy cost from the initial state. This part of the membrane therefore has an entropy cost $s_{eq} v_B t$. 
\end{enumerate}
For $t > (l+p)/v_B$ however, the above membrane is never minimal, as can be seen from considering the entropy of a membrane beginning at the edge of $B_1$ at time $t$ and ending on the left hand edge of $P$ at $t=0$ (shown in magenta in Figure~\ref{fig:islands13}). Such a membrane has a velocity $v_*(t) =(l + p)/t < v_B$, which contributes an entropy $s_{eq} {\cal E}(v_*(t)) t$. Note that for $\sat < \sign$ it is simple to show that convexity and monotonicity of the line tension ensure that $s_{eq} {\cal E}(v_*(t)) t$ is less than the contributions of all other surfaces considered above until the Page time, as we show in Appendix \ref{FurDet}. 
 \paragraph{} Taking into account that for all membranes there is an additional vertical piece of the membrane in the right hand subsystem contributing $s_{eq} v_E t$ to the entropy cost, we then find that the entropy of the bath region is given by the following expressions:
\begin{align}
    S(R, t) &= S_Q + 2\seq v_E t, && 0 \leq t \leq t_{\mathrm{sat}}, \nonumber \\
    &= \seq (p + l + v_E t),  && t_{\mathrm{sat}} \leq t \leq t_{\mathrm{signal}}, \nonumber \\
    &= \seq \mathcal{E}(v_*(t)) t + \seq v_E  t, &&  t_{\mathrm{signal}} \leq t \leq t_{P} + \Delta t/2, \nonumber \\
    &= 2S_{BH} + S_Q, && t \geq  t_{P} + \Delta t/2.
\end{align}
The entanglement entropy of the black hole is simpler to compute, receiving contributions from surfaces analogous to those in Figure~\ref{fig:islands12}, without the vertical cut shown near the reference system $Q$ at the left hand edge of diagram. We find
\begin{align}
    S(B,t) &= 2 \seq v_E t, && 0 \leq t \leq t_{P}, \nonumber \\
    &= 2S_{BH}, && t \geq t_{P}.
\end{align}
From which we can deduce the mutual information between the bath and the reference system is given by\footnote{Strictly speaking the behaviour for $t_{P} < t < t_{P} + \Delta t/2$ is given by $S_Q + \seq v_E t +  \seq \mathcal{E}(v_*(t)) t - 2 S_{BH}$. However for $t > t_P$ we have $v_* \approx 0$, giving the results stated. }
\begin{align}
    I(Q, R, t) &= 2 S_Q, && t \leq t_{\mathrm{sat}}, \nonumber \\
    &= 2S_Q + \seq (l- v_E t) , && t_{\mathrm{sat}} \leq t \leq  t_{\mathrm{signal}},  \nonumber\\
    &= S_Q + \seq (\mathcal{E}(v_*(t)) - v_E) t, && t_{\mathrm{signal}} \leq t \leq t_{P},  \nonumber\\
    &= S_Q + 2\seq v_E t - 2 S_{BH} , && t_{P} \leq t  \leq t_{P} + \Delta t/2, \nonumber \\
    &= 2S_Q,  && t \geq  t_{P} + \Delta t/2.
    \label{mutualfinal}
\end{align}
\paragraph{} To illustrate the above behaviour we can consider the von Neumann entropy $I_{vN}(Q, R, t)$ of a large $q$ random circuit with Poisson dynamics, for which we have ${\cal E}(v) = (1+ v^2)/2$ implying that $v_E = 1/2$ and $v_B = 1$. Then $\mathcal{E}(v_*(t)) t - v_E t = (l + p)^2/t$. The mutual information is then shown in Figure~\ref{fig:mutual} for a generic choice of parameters $l=10, p =15,  t_P = 400$. We see that the approximately half the mutual information initially inside the bath is rapidly transferred to the black hole after $t_{\mathrm{sat}}$. The initial decay of $I(Q, R, t)$ is linear for $t_{\mathrm{sat}} < t <  t_{\mathrm{signal}}$ but then becomes a power law asymptoting to $S_Q$ for $t > t_{\mathrm{signal}}$. In particular, for timescales $t \gg t_{\mathrm{signal}}$ approximately half the information has been transferred to the black hole. This information however is then rapidly returned to the radiation around the Page time, over a timescale of order $\Delta t$. 
\paragraph{}Similar behaviour can be seen if one computes the second Renyi entropy $I_2(Q,R,t)$ for a large-$q$ brickwork random circuit, noting that in this case ${\cal E}(v) \to 1$, $v_E \to 1$, $v_B \to 1$ as $q \to \infty$. The results are qualitatively similar to those of the von Neumann entropy for the Poisson model, although in this case the flatness of the line tension means that the mutual information is flat $I_2(Q, R,  t) = S_Q$ for $t_{\mathrm{signal}} < t < t_{P}$. Our results for $I_2(Q,R,t)$ agree precisely with those presented in~\cite{Liu_2021}, which studied the Hayden-Preskill protocol in a related model using void formation, and are illustrated in the right hand panel of Figure~\ref{fig:mutual} for  $l=10, p =15,  t_P = 400$.
\subsubsection{$\sign < \sat$}
\paragraph{} We now consider the case $\sign < \sat$,  i.e. $(l+p)/v_B <  l/v_E$. In this case, the orange membrane above is never minimal since its entropy only become less than the red membrane for $t > \sat$, by which time we are now guaranteed that the magenta membrane exists and has a smaller entropy. The behaviour of the mutual information is now determined by a competition between the red and magenta membranes. For a generic line tension ${\cal E}(v)$ we then find the behaviour 
\begin{align}
    S(R, t) &= S_Q + 2 \seq v_E t,  && 0 \leq t \leq t^*, \nonumber \\
    &= \seq \mathcal{E}(v_*(t))  t + \seq v_E  t , && t^* \leq t \leq t_P + \Delta t/2, \nonumber \\
    &= 2S_{BH} + S_Q, && t \geq t_P + \Delta t/2,
\end{align}
and
\begin{align}
    I(Q, R, t) &= 2S_Q, && 0 \leq t \leq t^*, \nonumber\\
    &= S_Q +  s_{eq} (\mathcal{E}(v_*(t)) - v_E) t, && t^* \leq t \leq  t_P, \nonumber \\
    &= S_Q +  2 \seq v_E t - 2 S_{BH}, &&   t_P \leq t \leq t_P + \Delta t/2  \nonumber \\
    &= 2 S_Q, && t>t_P + \Delta t/2.
\end{align}
were $t^*$ is the timescale at which the red and magenta membranes exchange dominance. This is the smallest timescale such that  
\begin{equation}
 s_{eq} (\mathcal{E}(v_*(t)) - v_E) t \leq S_Q, 
\label{crossover}
\end{equation}
For the special case of the von Neumann entropy of a large-$q$ random circuit with Poisson dynamics, i.e. with line tension $\mathcal{E}(v) = (1 + v^2)/2$, one has $t^* = (l+p)^2/2p$ and the mutual information is given by 
\begin{align}
    I_{vN}(Q, R, t) &= 2S_Q, && 0 \leq t \leq t^*, \nonumber \\
    &= S_Q + \seq (l+ p)^2/2t, && t^* \leq t \leq  t_P, \nonumber\\
    &= S_Q +  \seq t - 2 S_{BH} , &&   t_P \leq t \leq t_P + \Delta t/2, \nonumber \\
    &= 2 S_Q, && t \geq t_P + \Delta t/2.
\end{align}
We therefore see that there is a non-trivial crossover between the magenta and red surfaces at the intermediate timescale $t^*$, which is responsible for the transfer of information to the black hole. That information is then recovered from the black hole rapidly around the Page time. Qualitatively similar behaviour is obtained for $S_2$ in the brickwork quantum circuit using the line tension~\eqref{eps2}. 

\paragraph{} The above behaviour is generic to any line tension ${\cal E}(v)$ for which there is a non-trivial cross-over between the red and magenta membranes at time $t^* \ll  t_P$. It is simple to check, that~\eqref{crossover} will be satisfied at sufficiently large $t$ provided that ${\cal E}(v)$ is analytic at $v=0$. This can be seen from the fact that with these assumptions then at sufficiently large $t$ (i.e. sufficiently small $v_*(t)$) the line tension can be Taylor expanded as
\begin{equation}
{\cal E}(v_*(t))  \approx v_E + {\cal E}''(0)(l+p)^2/2 t^2
\end{equation}
where we have used that since ${\cal E}(v)$ is even we must have ${\cal E}'(0)=0$ provided the line tension is analytic at $v=0$. However it is interesting to note that such a crossover would not be seen if the bath was described by a non-analytic line tension of the form $\mathcal{E}(v) = v_E + |v|(1 -  v_E/v_B)$, as we show in Appendix~\ref{FurDet}. Whilst we are not aware of any microscopic system described by such a line tension\footnote{A system with such a line tension also saturates various bounds on entanglement growth proposed in~\cite{Mezei_2017}.} it is noteworthy that in this case we find that for $\sat > \sign$ the red membrane remains minimal until the Page time, and as a result the information encoded in the bath is never transferred to the black hole.

\section{Discussion}
\label{sec:discussion}

\paragraph{} In this paper we have seen that the entanglement membrane provides a powerful framework through which to understand entanglement dynamics in models of black hole information built out of chaotic many-body quantum systems. Whilst the models we studied were highly simplified descriptions of black hole dynamics, we were able to reproduce the expected features of both the Page curve and the Hayden-Preskill protocol. We emphasise that our results required very little input as to the underlying nature of the microscopic physics, aside from the assumptions of a system whose local chaotic unitary dynamics could be describing using the entanglement membrane. The results we obtained for the eternal black hole in Section~\ref{sec:eternal} relied essentially only on this assumption and a model of a black hole as a region of many, dense, degrees of freedom. As such our results suggest these aspects of black hole information dynamics are a consequence of generic features of entanglement dynamics in chaotic many-body quantum systems. 

\paragraph{} Furthermore, one key motivation for studying information dynamics in these models using the entanglement membrane prescription is that it provides an approach to studying the Page curve in these models that is reminiscent of gravitational studies of entanglement dynamics using (quantum) extremal surfaces~\cite{Ryu_2006, Hubeny_2007, Engelhardt_2015, penington2020entanglement, Almheiri_2019, Almheiri_2020}. Whilst we leave a detailed comparison to future work, we wish to close this paper with some preliminary observations about possible connections between the work in this paper and gravitational approaches to studying entanglement dynamics.

\paragraph{} There already exists a well established connection between the entanglement membrane and gravitational computations of entanglement entropies in holographic quantum field theories using classical extremal surfaces\cite{Mezei_2018, Mezei_2020}. Specifically, the entanglement dynamics in the boundary of a holographic quantum field theory admits a coarse-grained description in terms of an entanglement membrane. In this context, the entanglement membrane emerges as the projection of the classical bulk HRT surface into the boundary theory. It has been established that in certain models of black hole evaporation the Page curve can be derived simply from considering classical HRT surfaces in a ``doubly holographic'' spacetime~\cite{Almheiri_2020, chen2020evaporating, chen2020quantum1, chen2020quantum2, Geng:2020qvw, Geng:2020fxl}. It would be interesting to study the projections of such surfaces into the boundary, which could allow for a direct relationship between the entanglement membrane studied here and the islands prescription in gravity. Furthermore, currently the entanglement membrane has only been derived from holography in the case where the bulk theory is classical. It would be interesting to understand if one can provide a similar derivation of the entanglement membrane in the case where the gravitational extremal surface problem includes the effects of quantum corrections~\cite{Engelhardt_2015}.

\paragraph{} Moreover, it would be interesting to explore the relationship between replica wormholes and permutation degrees of freedom that are naturally connected to the entanglement membrane proposal~\cite{Vardhan_2021, PhysRevX.10.031066, Zhou_2019}. In particular, in large-$q$ random quantum circuits it has been established that the entanglement membrane can be understood as a domain wall between regions of the circuit in which gates in different replica copies are identified according to different permutations of the replicas~\cite{PhysRevX.10.031066, Zhou_2019}. Similar permutation degrees of freedom have also been identified as governing the entanglement entropies of thermalized states in general many-body systems including quantum field theories~\cite{Vardhan_2021}. The connection between the entanglement membrane and such permutation degrees of freedom is most easily illustrated by considering the computation of the second Renyi entropy 
\begin{equation}
S_2(A, t) = -\log\mathrm{Tr}_A(\rho^2_{A}(t)), \hspace{4.0cm} \rho_{A}(t) = \mathrm{Tr}_{\bar A} \ket{\psi(t)}\bra{\psi(t)}, 
\label{trace}
\end{equation}
where $A$ is a subregion of the systems discussed in Sections~\ref{sec:evap0} and~\ref{sec:eternal} and ${\bar A}$ its complement. 

\paragraph{} In order to compute such a Renyi entropy we require two copies of the density matrix, which can be thought of as a state in a four copy Hilbert space with time evolution generated by $U_1 \otimes U_1^{\dagger} \otimes U_2 \otimes U_2^{\dagger}$, with $U_i, U_i^{\dagger}$ built from independently drawn random unitaries (gates). Then the insight of~\cite{PhysRevX.10.031066, Zhou_2019} is that after averaging over random unitaries one obtains a description in terms of a permutation degree of freedom $\sigma(t',x)$, that describes whether at each location in spacetime one identifies the gates associated to different replicas pairwise through either $U_1 = U_1^{\dagger}, U_2 = U_2^{\dagger}$ or via $U_1 = U_2^{\dagger}, U_2 = U_1^{\dagger}$. We will refer to the first configuration as identifying the gates using the identity permutation $\sigma = e$ and the second via a swap configuration $\sigma = \eta$. The entanglement membrane can then be understood as representing a domain wall configuration across which the identification of unitaries changes from $e$ to $\eta$ (or vice versa). 

\begin{figure}
\begin{center}
\includegraphics[width=60mm]{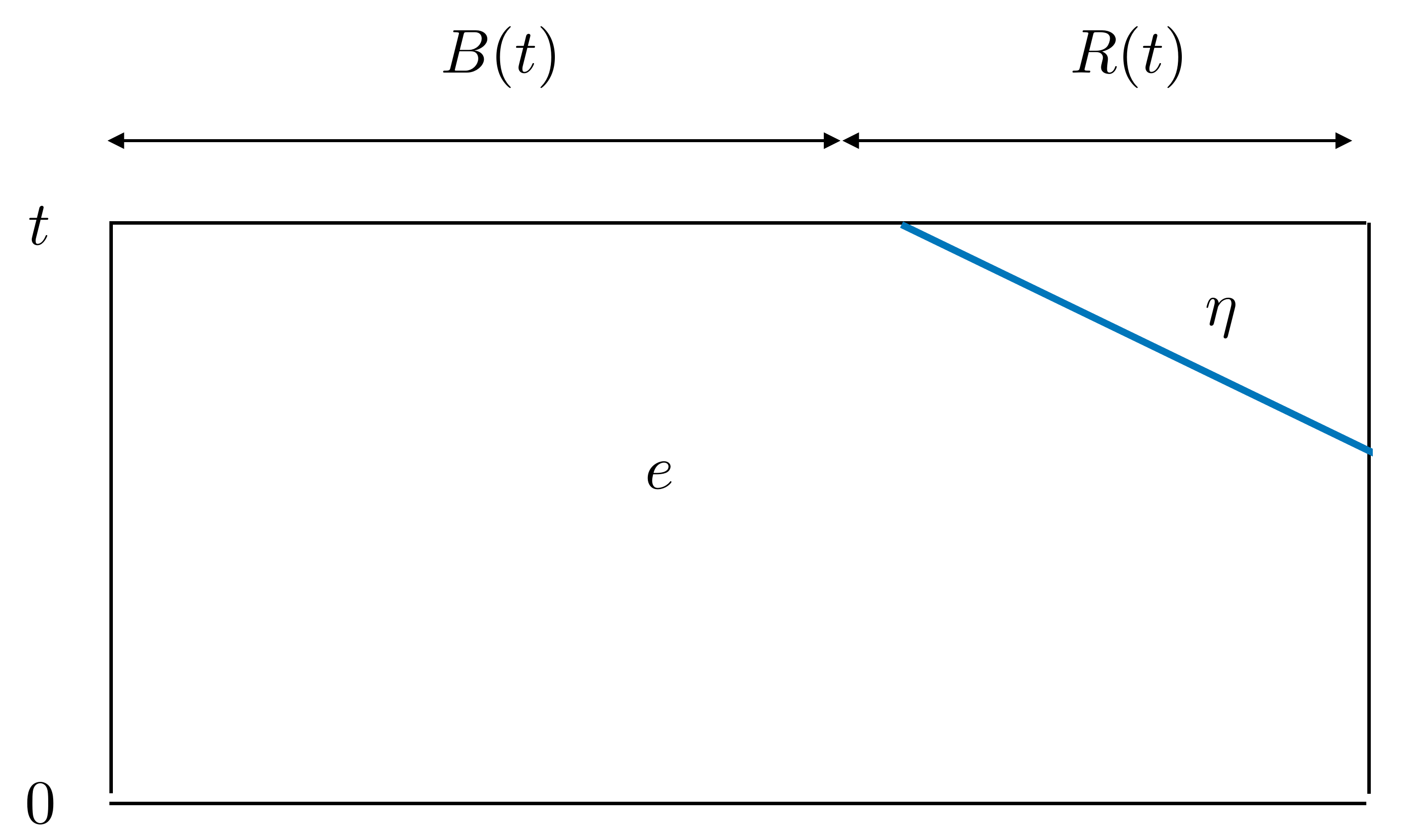}
\includegraphics[width=60mm]{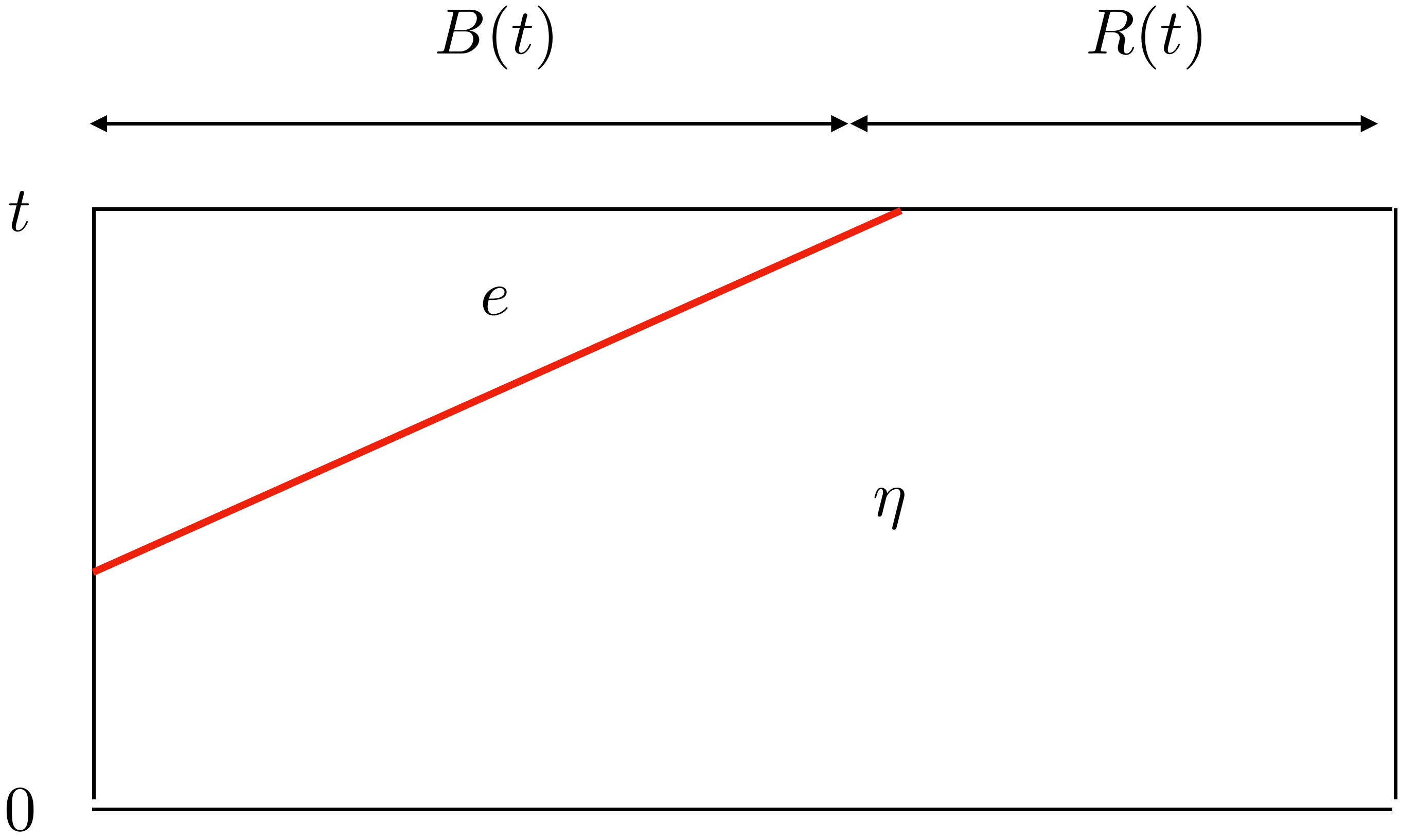}
\caption{Permutations describing the identification between different replicas in the computation of $S_2(R, t)$ in our model of an evaporating black hole.}
\label{fig:permevap}
\end{center}
\end{figure}

\begin{figure}
\begin{center}
\includegraphics[width=70mm]{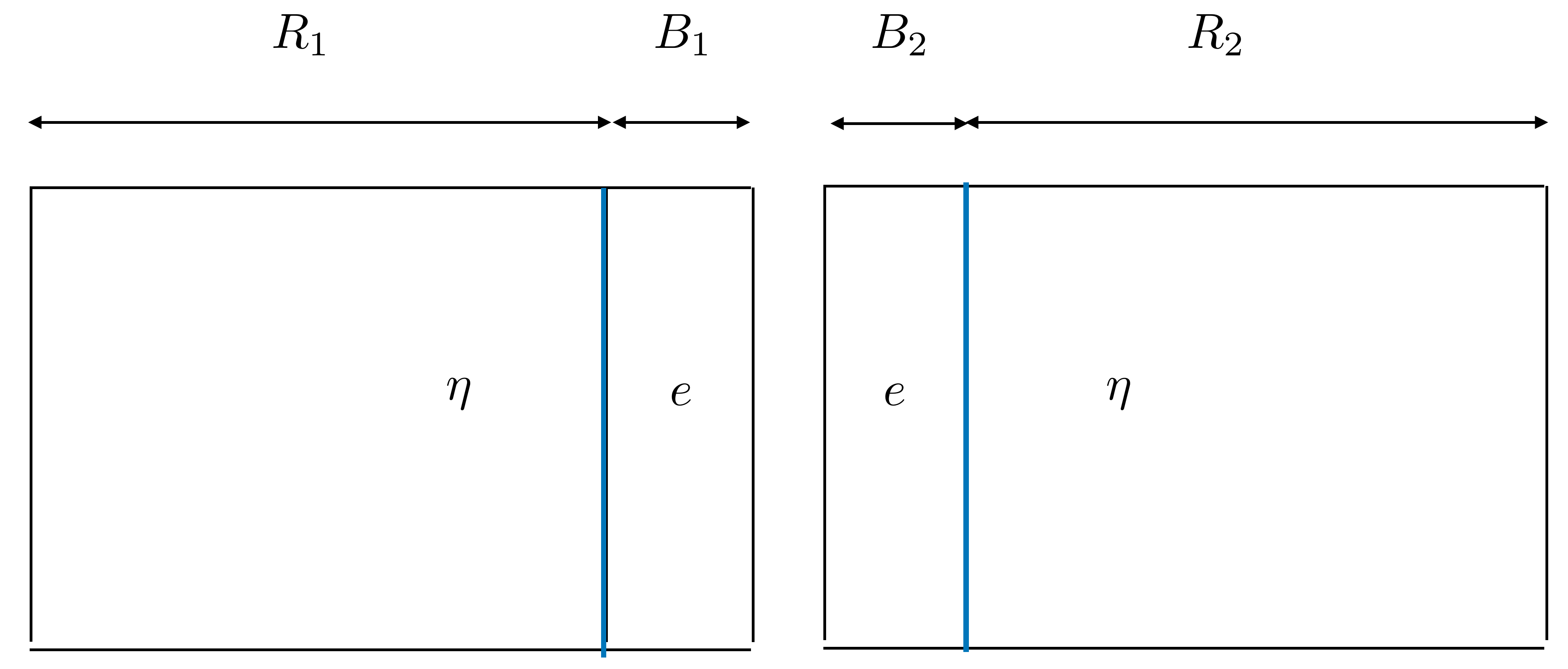}
\includegraphics[width=70mm]{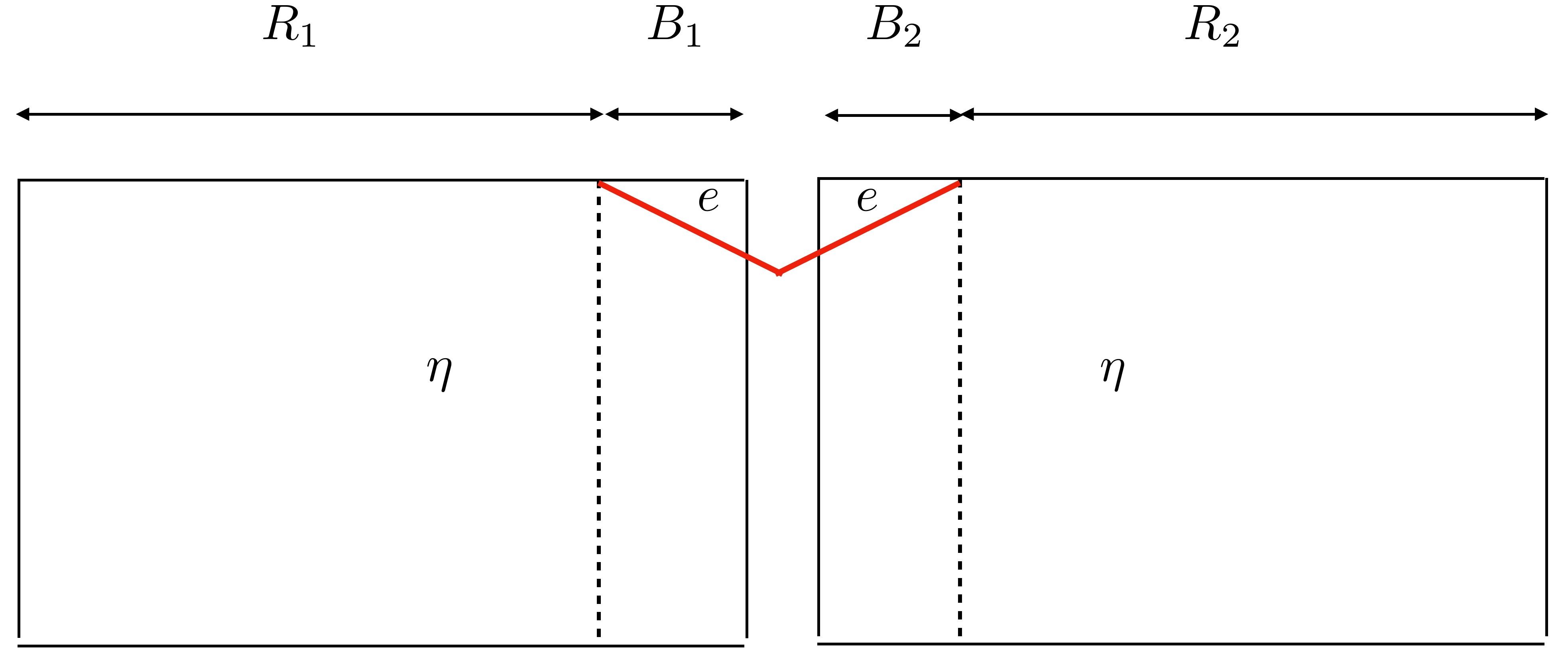}
\caption{Permutations describing the identification between different replicas in the computation of $S_2(R, t)$ in our model of an eternal black hole.}
\label{fig:permeternal}
\end{center}
\end{figure}

\paragraph{} It is instructive to illustrate the identification of the replicas in the model of the evaporating black hole discussed in Section~\ref{sec:evap0}, where we take $A = R(t)$ and $\bar{A} = B(t)$. As discussed in~\cite{PhysRevX.10.031066, Liu_2021} the contractions implied by the trace in \eqref{trace} imply boundary conditions at $t' = t$ on the permutation $\sigma$ such that $\sigma(t, x) = e$ in $B(t)$ and $\sigma(t, x) = \eta$ in $R(t)$. The permutations at times $t'< t$ are then determined dynamically by the entanglement membrane prescription. In particular, the permutations associated with the two diagrams contributing to the Page curve are shown in Figure~\ref{fig:permevap} (see also Figure~\ref{fig:permeternal} for the analogous diagrams in our model of an eternal black hole). We note that the left hand diagram, associated to the early time behaviour of the Page curve, corresponds to identifying replicas at small $t'$ according to $\sigma = e$, whilst the late time behaviour is related to identifying them at small $t'$ as $\sigma = \eta$.

\paragraph{} The fact that the Page curve arises from a transition between different ways of connecting multiple replicas is highly reminiscent of gravitational computations of the second Renyi entropy, in which there is a transition around the Page time between different ways of connecting the interiors of black holes associated to each replica (so-called `replica wormhole' geometries). Similar permutation degrees of freedom have been shown to govern the entanglement entropies of thermalized states in generic many-body quantum systems, and in this context a precise connection to replica wormholes has been established in~\cite{Vardhan_2021}. In our above discussion of an evaporating black hole both the radiation and black hole systems are essentially thermalized at all times, and the permutation degrees of freedom in Figure~\ref{fig:permevap} can be identified with those obtained by the prescription of~\cite{Vardhan_2021}\footnote{Whilst we restricted to considering membranes inside the butterfly cone, the membranes in Fig.~\ref{fig:permevap} can be freely deformed outside the butterfly-cone into horizontal membranes~\cite{jonay2018coarsegrained}. The permutations in Fig.~\ref{fig:permevap} then become $e$ and $\eta$ respectively everywhere the bulk, which can be interpreted as the terms in (2.34) of~\cite{Vardhan_2021} for $S_2$. We are grateful to Shreya Vardhan for discussions on this point.}.

\paragraph{}However the entanglement membrane prescription is more general in the sense that it can be applied to non-thermal states, which was relevant for our discussion of the Page curve in the eternal black hole. For instance whilst~\cite{Vardhan_2021} was able to apply techniques developed for thermalized states to obtain the Page curve from permutation degrees of freedom in a model of an eternal black hole, this could only be done for the special case where the bath regions $R_1,R_2$ were built from a many-body system with `maximally efficient thermalization' corresponding to $v_E = v_B = 1$. For the more generic models studied here in Section~\ref{sec:eternal}, with an arbitrary line tension ${\cal E}(v)$ in the bath regions, the techniques of~\cite{Vardhan_2021} cannot be applied at times $t < t_P$. It would therefore be extremely interesting to perform a more detailed study of the relationship between permutation degrees of freedom and gravitational computations of replica wormholes in non-thermalized states such as this in the future\footnote{Given that the entanglement membrane is formulated in a Lorentzian setting the most natural starting point for such a comparison may be provided by~\cite{Marolf_2021a, Marolf_2021b}.}. 

\acknowledgments
We are grateful to Hong Liu, M\'{a}rk Mezei, Adam Nahum, Shreya Vardhan and Curt von Keyserlingk for very helpful discussions. AT acknowledges support from UK EPSRC (EP/SO23607/1). MB acknowledges support from UK Research and Innovation (UKRI) under the UK government's Horizon Europe guarantee (EP/Y00468X/1).

\appendix
\addcontentsline{toc}{section}{Appendices}
\renewcommand{\thesection}{\Alph{section}}

\section{Mutual information of bath regions in eternal black hole}
\label{app:mutual}
\paragraph{} In this appendix we wish to consider the mutual information $I(R_1, R_2, t)$ between the two bath regions in our model of an eternal black hole discussion in Section~\ref{sec:eternal}. In particular, we note that \cite{Liu_2021} studied this (second Renyi) mutual information in a similar dynamical model of an eternal black hole using void formation, and found that the generation of mutual information was sensitive to the entanglement structure of initial state of the bath regions $R_1, R_2$. Here we will see how the same physics arises using the entanglement membrane.

\paragraph{} In particular, recall that the mutual information is defined as 
\begin{equation}
I(R_1, R_2, t) = S(R_1, t) + S(R_2, t) - S(R_1 \cup R_2, t). 
\label{mutualbath}
\end{equation}
As in the main text we will take the black hole regions $B_1$ and $B_2$ to be in a thermofield double state (or maximally entangled) with entropy density $\tilde{s}_{eq}$. We now wish to compute~\eqref{mutualbath} for two choices of initial state in the bath region: (i) When $R_1$, $R_2$ are in a product state at $t=0$; (ii) When $R_1$ and $R_2$ are in a thermofield double state with entropy density $s_{eq}$.
\paragraph{} For the case $R_1$, $R_2$ are initially in a product state, we have already determined $S(R_1 \cup R_2, t)$ in equation \eqref{eternalBHentropy}. Since by symmetry we have $S(R_1, t) = S(R_2,  t)$ we then only need to compute $S(R_1, t)$. The relevant entanglement membranes are
\begin{figure}
\begin{center}
\includegraphics[width=120mm]{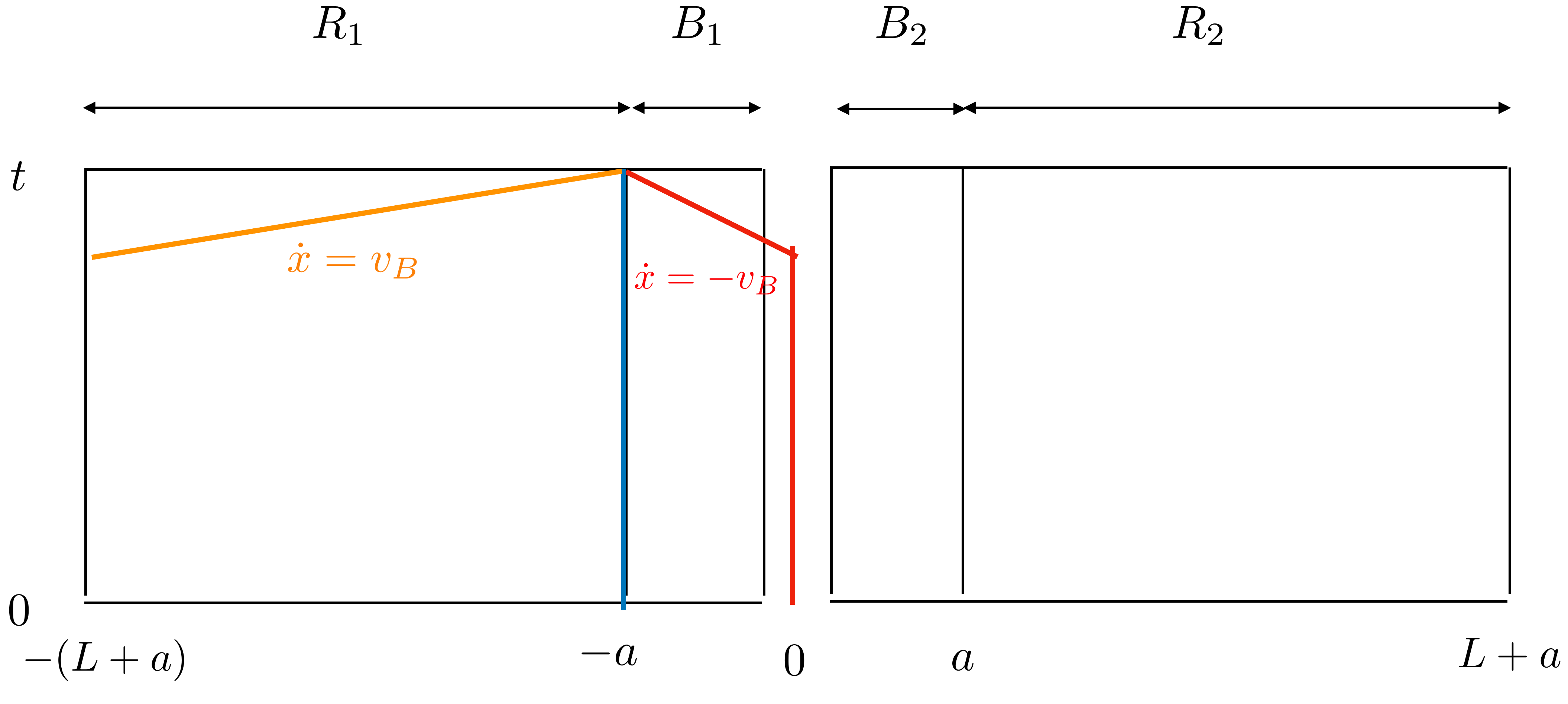}
\caption{Candidate minimal surfaces for $S(R_1,  t)$.}
\label{fig:islands4}
\end{center}
\end{figure}
\begin{enumerate}
\item A vertical cut down from $x \approx -a$ to the initial state. This has entropy $s_{eq} v_E t + S(-a,0) = s_{eq} v_E t$. 
\item A horizontal cut from $x=-a$ to $x=0$, joined to a vertical cut down to the initial state through the gap in the space-time slab at $x=0$. This has entropy $S_{BH} + S(0,0) = 2 S_{BH}$ where $S_{BH} = \tilde{s}_{eq} a$ as in the main text. 
\end{enumerate}
Taking the minimal membrane one finds
\begin{align}
S(R_1,t) &= s_{eq} v_E t, && t \leq 2 t_P, \nonumber \\
&= 2 S_{BH}, && t \geq 2 t_P,
\end{align}
with $t_P = S_{BH}/(s_{eq} v_E)$ the Page time of the eternal black hole. The mutual information $I(R_1, R_2, t)$ is then given by:
\begin{align}
I(R_1, R_2, t) &= 0,  &&  t \leq t_P, \nonumber \\
&=  2 s_{eq} v_E t - 2 S_{BH}, && t_P \leq t  \leq 2 t_P, \nonumber \\
&= 2 S_{BH}, && t \geq 2 t_P.
\label{mutualbath1}
\end{align}
\paragraph{} If we instead take $R_1$ and $R_2$ to initially be in a thermofield double state, the computation of $S(R_1 \cup R_2, t)$ in equation \eqref{eternalBHentropy} remains unchanged, as the initial state entropy term continues to vanish for both the surfaces contributing to~\eqref{eternalBHentropy}. In contrast, the entropy cost of the surfaces relevant for  $S(R_1, t)$ shown in Figure~\ref{fig:islands4} are now modified to 
\begin{enumerate}
\item $s_{eq} v_E t + S(-a,0) = s_{eq} v_E t + s_{eq} L$ for the blue surface.
\item $S_{BH} + S(0,0) = 2 S_{BH} + s_{eq} L$ for the red surface. 
\end{enumerate}
\paragraph{} For this initial state, the entropy of both these surfaces is now larger than that of arising due to a membrane from $x=-a$ to the left hand edge of the system with $\dot{x} = v_B$ (the yellow surface in Fig.~\ref{fig:islands4}). This has entropy $s_{eq} L$ and remains the minimal membrane for all $t$. We therefore have a constant entropy $S(R_1, t) = s_{eq} L$, which results in the mutual information
\begin{align}
I(R_1, R_2, t) &= 2 s_{eq} L- 2 s_{eq} v_E t,  && t \leq t_P, \nonumber \\
&= 2 s_{eq} L - 2 S_{BH}, && t \geq  t_P. 
\label{mutualbath2}
\end{align}
The results~\eqref{mutualbath1} and~\eqref{mutualbath2} precisely match those in Section III of \cite{Liu_2021} for the appropriate entanglement velocity $v_E =1$. 

\section{Minimal membranes for generic line tensions}\label{FurDet}

\paragraph{} In this section we will justify various claims we made in our discussions of the Hayden Preskill protocol in our model of the eternal black hole in Section~\ref{InfOut}. 

\paragraph{} Firstly, we claimed that when $\sat < \sign$ the behaviour of the entropy $S(R,t)$ and mutual information $I(Q, R, t)$ were qualitatively the same for all forms of the line tension $\mathcal{E} (v)$. In particular, we claimed that when $t > \sign$ the entropy of the magenta membrane with $v_*(t) = (l+p)/t$ is always less than the orange or red membranes. 
To show this we recall from the equations in~\eqref{equality} and~\eqref{inequality} that the line tension satisfies ${\cal E}(v) \leq {\cal E}(v_B) = v_B$, where generically the equality only holds for $|v|=v_B$. We therefore have that in generic models $s_{eq} {\cal E}(v_*(t)) t < s_{eq} v_B t $, implying that the entropy of the magenta membrane is less than the orange membrane. Furthermore, to compare the entropy of the magenta membrane to the vertical red membrane we note that using the third inequality in~\eqref{inequality} we have
\begin{equation}
s_{eq} {\cal E}(v_*(t)) t \leq s_{eq} v_E t + s_{eq} (l + p)\bigg(1 - \frac{v_E}{v_B}\bigg) = s_{eq} v_E t + S_Q + s_{eq} v_E(\sat - \sign). 
\label{magred}
\end{equation}
Hence for $\sat < \sign$ we have that $s_{eq} {\cal E}(v_*(t)) t < s_{eq} v_E t + S_Q$, implying that the entropy of the magenta membrane is less than the red membrane. 

\paragraph{} Secondly, we noted for $\sat > \sign$ that if one considers a non-analytic form of line tension the line tension $\mathcal{E}(v) = v_E + |v|(1 -  v_E/v_B)$ then for $\sat > \sign$ information initially encoded in the bath is never transferred to the black hole. This follows from the fact that for such a line tension we have that the inequality in \eqref{magred} is saturated, implying $s_{eq} {\cal E}(v_*(t)) t > s_{eq} v_E + S_Q$ for $\sat > \sign$. As a result the red membrane remains minimal until the Page time, and we find 
\begin{align*}
    S(R , t) &= S_Q + 2v_E \seq t, && t \leq t_P, \\
    &= 2S_{BH} + S_Q, && t \geq t_P,
\end{align*}
which implies that the mutual information $I(Q, R, t) = 2S_Q$ for all time.

\paragraph{} Finally, throughout our discussion we argued that before the Page time the minimal membrane is always one of the surfaces shown in the diagrams in Section~\ref{InfOut} (i.e. the red, orange or magenta curves). It is usually straightforward to check that any other membrane would result in a larger entropy cost than at least one of the membranes discussed in the main text. However, the case where the membrane ends in the region $P$ requires a more careful justification, due to the non-trivial entropy profile of the initial state in this region. Therefore, suppose we consider a membrane of velocity $v_0$ starting from the left hand edge of $B_1$ and intersecting $t=0$ somewhere in the region $P$. This requires $l/t \leq v_0 \leq (l+p)/t$. The entropy of such a membrane is
\begin{equation}
s_{eq} {\cal E}(v_0) t + s_{eq}(l + p - v_0 t) = s_{eq}(l +p) + s_{eq}({\cal E}(v_0) - v_0) t.
\end{equation}
For a generic line tension satisfying ${\cal E}'(v_B) =1$ and ${\cal E}''(v) \geq 0$  (see equations~\eqref{equality} and~\eqref{inequality}) we have that ${\cal E}'(v) < 1$ for $v < v_B$. As such $({\cal E}(v_0) - v_0)$ is a monotonically decreasing function, and so the smallest entropy cost is obtained for a membrane with the largest possible choice of $v_0$. For we therefore find that the minimal membrane intersecting $P$ at $t=0$ has velocity $v_B$ for $t < \sign$ and velocity $v_*(t)$ for $t> \sign$ as claimed.

\bibliographystyle{JHEP}
\bibliography{bib.bib}

\end{document}